\documentstyle[12pt,bezier,epsf]{article}

\makeatletter
\def\eqnarray{\stepcounter{equation}\let\@currentlabel=\theequation
\global\@eqnswtrue
\global\@eqcnt\z@\tabskip\@centering\let\\=\@eqncr
$$\halign to \displaywidth\bgroup\@eqnsel\hskip\@centering
  $\displaystyle\tabskip\z@{##}$&\global\@eqcnt\@ne
  \hfil$\displaystyle{\hbox{}##\hbox{}}$\hfil
  &\global\@eqcnt\tw@ $\displaystyle\tabskip\z@
  {##}$\hfil\tabskip\@centering&\llap{##}\tabskip\z@\cr}
\def\@sect#1#2#3#4#5#6[#7]#8{\ifnum #2>\c@secnumdepth
    \def\@svsec{}\else
    \refstepcounter{#1}\edef\@svsec{\csname the#1\endcsname.\hskip 1em }\fi
    \@tempskipa #5\relax
    \ifdim \@tempskipa>\z@
    \begingroup #6\relax
    \@hangfrom{\hskip #3\relax\@svsec}{\interlinepenalty \@M #8\par}
    \endgroup
    \csname #1mark\endcsname{#7}\addcontentsline
    {toc}{#1}{\ifnum #2>\c@secnumdepth \else
     \protect\numberline{\csname the#1\endcsname}\fi
           #7}\else
    \def\@svsechd{#6\hskip #3\@svsec #8\csname #1mark\endcsname
          {#7}\addcontentsline
          {toc}{#1}{\ifnum #2>\c@secnumdepth \else
     \protect\numberline{\csname the#1\endcsname}\fi
           #7}}\fi
     \@xsect{#5}}
\def\label#1{\@bsphack\if@filesw {\let\thepage\relax
   \xdef\@gtempa{\write\@auxout{\string
   \newlabel{#1}{{\thesection.\@currentlabel}{\thepage}}}}}\@gtempa
   \if@nobreak \ifvmode\nobreak\fi\fi\fi\@esphack}
\def\@eqnnum{(\thesection.\theequation)}
\def\section{\setcounter{equation}{0} \@startsection {section}{1}{\z@}{-3.5ex
   plus -1ex minus -.2ex}{2.3ex plus .2ex}{\Large\bf}}
\newcount\@minsofar
\newcount\@min
\newcount\@cite@temp
\def\@citex[#1]#2{%
\if@filesw \immediate \write \@auxout {\string \citation {#2}}\fi
\@tempcntb\m@ne \let\@h@ld\relax \def\@citea{}%
\@min\m@ne%
\@cite{%
  \@for \@citeb:=#2\do {\@ifundefined {b@\@citeb}%
    {\@h@ld\@citea\@tempcntb\m@ne{\bf ?}%
    \@warning {Citation `\@citeb ' on page \thepage \space undefined}}%
{\@minsofar\z@ \@for \@scan@cites:=#2\do {%
  \@ifundefined{b@\@scan@cites}%
    {\@cite@temp\m@ne}
    {\@cite@temp\number\csname b@\@scan@cites \endcsname \relax}%
\ifnum\@cite@temp > \@min
    \ifnum\@minsofar = \z@
      \@minsofar\number\@cite@temp
      \edef\@scan@copy{\@scan@cites}\else
    \ifnum\@cite@temp < \@minsofar
      \@minsofar\number\@cite@temp
      \edef\@scan@copy{\@scan@cites}\fi\fi\fi}\@tempcnta\@min
  \ifnum\@minsofar > \z@ 
    \advance\@tempcnta\@ne
    \@min\@minsofar
    \ifnum\@tempcnta=\@minsofar 
      \ifx\@h@ld\relax
        \edef \@h@ld{\@citea\csname b@\@scan@copy\endcsname}%
    \else \edef\@h@ld{\ifmmode{-}\else--\fi\csname b@\@scan@copy\endcsname}%
      \fi
    \else \@h@ld\@citea\csname b@\@scan@copy\endcsname
          \let\@h@ld\relax
  \fi 
\fi}%
\def\@citea{,\penalty\@highpenalty\,}}\@h@ld}{#1}}
\def\appendixname{Appendix}
\def\appendix{\par
  \def\pre@section{\appendixname{}}
  \setcounter{section}{1}
  \@addtoreset{equation}{section}
  \def\thesection{\Alph{section}}
  \def\theequation{\arabic{equation}}}

\def\appendix{\par
  \def\pre@section{\appendixname{}}
  \setcounter{section}{1}
  \@addtoreset{equation}{section}
  \def\thesection{\Alph{section}}
  \def\theequation{\arabic{equation}}}

\makeatother

\begin{document}
\addtolength{\unitlength}{-0.5\unitlength}

\def\b{\beta}
\def\d{\delta}
\def\g{\gamma}
\def\a{\alpha}
\def\s{\sigma}
\def\t{\tau}
\def\th{\theta}
\def\l{\lambda}
\def\la{\lambda}
\def\e{\epsilon}
\def\r{\rho}
\def\wid{\widehat}
\def\ds{\displaystyle}
\def\be{\begin{equation}}
\def\ee{\end{equation}}
\def\beq{\begin{eqnarray}}
\def\eeq{\end{eqnarray}}
\def\ov{\overline}
\def\om{\omega}
\vspace{2cm}

\newsavebox{\neara}\savebox{\neara}(30,30){\begin{picture}(30,30)
\thicklines\put(10,20){\line(1,-1){10}}\put(10,20){\line(2,1){20}}
\put(20,10){\line(1,2){10}}\end{picture}}

\newsavebox{\nwara}\savebox{\nwara}(30,30){\begin{picture}(30,30)
\thicklines\put(10,10){\line(1,1){10}}\put(10,10){\line(-1,2){10}}
\put(20,20){\line(-2,1){20}}\end{picture}}

\newsavebox{\swara}\savebox{\swara}(30,30){\begin{picture}(30,30)
\thicklines\put(10,20){\line(1,-1){10}}\put(0,0){\line(2,1){20}}
\put(0,0){\line(1,2){10}}\end{picture}}

\newsavebox{\seara}\savebox{\seara}(30,30){\begin{picture}(30,30)
\thicklines\put(10,10){\line(1,1){10}}\put(10,10){\line(2,-1){20}}
\put(20,20){\line(1,-2){10}}\end{picture}}

\newsavebox{\nara}\savebox{\nara}(30,30){\begin{picture}(30,30)
\thicklines\put(5,10){\line(1,0){20}}\put(5,10){\line(1,2){10}}
\put(15,30){\line(1,-2){10}}\end{picture}}

\newsavebox{\wara}\savebox{\wara}(30,30){\begin{picture}(30,30)
\thicklines\put(0,15){\line(2,1){20}}\put(0,15){\line(2,-1){20}}
\end{picture}}

\newsavebox{\sara}\savebox{\sara}(30,30){\begin{picture}(30,30)
\thicklines\put(15,0){\line(1,2){10}}\put(15,0){\line(-1,2){10}}
\end{picture}}

\newsavebox{\eara}\savebox{\eara}(30,30){\begin{picture}(30,30)
\thicklines\put(10,5){\line(2,1){20}}
\put(10,25){\line(2,-1){20}}\end{picture}}

\newsavebox{\near}\savebox{\near}(30,30){\begin{picture}(30,30)
\thicklines\put(10,20){\line(1,-1){10}}\put(10,20){\line(2,1){20}}
\put(20,10){\line(1,2){10}}\end{picture}}

\newsavebox{\nwar}\savebox{\nwar}(30,30){\begin{picture}(30,30)
\thicklines\put(10,10){\line(1,1){10}}\put(10,10){\line(-1,2){10}}
\put(20,20){\line(-2,1){20}}\end{picture}}

\newsavebox{\swar}\savebox{\swar}(30,30){\begin{picture}(30,30)
\thicklines\put(10,20){\line(1,-1){10}}\put(0,0){\line(2,1){20}}
\put(0,0){\line(1,2){10}}\end{picture}}

\newsavebox{\sear}\savebox{\sear}(30,30){\begin{picture}(30,30)
\thicklines\put(10,10){\line(1,1){10}}\put(10,10){\line(2,-1){20}}
\put(20,20){\line(1,-2){10}}\end{picture}}

\newsavebox{\nar}\savebox{\nar}(30,30){\begin{picture}(30,30)
\thicklines\put(5,10){\line(1,0){20}}\put(5,10){\line(1,2){10}}
\put(15,30){\line(1,-2){10}}\end{picture}}

\newsavebox{\war}\savebox{\war}(30,30){\begin{picture}(30,30)
\thicklines\put(0,15){\line(2,1){20}}\put(0,15){\line(2,-1){20}}
\put(20,5){\line(0,1){20}}\end{picture}}

\newsavebox{\sar}\savebox{\sar}(30,30){\begin{picture}(30,30)
\thicklines\put(15,0){\line(1,2){10}}\put(15,0){\line(-1,2){10}}
\put(5,20){\line(1,0){20}}\end{picture}}

\newsavebox{\ear}\savebox{\ear}(30,30){\begin{picture}(30,30)
\thicklines\put(10,5){\line(0,1){20}}\put(10,5){\line(2,1){20}}
\put(10,25){\line(2,-1){20}}\end{picture}}

\newsavebox{\mybox}\savebox{\mybox}(225,225){
\begin{picture}(225,225)
\thicklines
\put(75,0){\line(1,0){150}}\put(75,0){\line(0,1){150}}
\put(75,150){\line(1,0){150}}\put(225,0){\line(0,1){150}}
\put(75,0){\line(-1,1){75}}\put(75,150){\line(-1,1){75}}
\put(225,150){\line(-1,1){75}}\put(0,75){\line(0,1){150}}
\put(0,225){\line(1,0){150}}\multiput(0,75)(0,150){2}{\circle*{10}}
\multiput(0,75)(150,0){2}{\circle*{10}}
\multiput(75,0)(0,150){2}{\circle*{10}}
\multiput(75,0)(150,0){2}{\circle*{10}}
\multiput(225,150)(-75,75){2}{\circle*{10}}
\multiput(0,75)(25,0){6}{\line(1,0){15}}
\multiput(150,75)(0,25){6}{\line(0,1){15}}
\multiput(225,0)(-25,25){3}{\line(-1,1){20}}
\end{picture}}

\newsavebox{\stars}\savebox{\stars}(350,350){
\begin{picture}(350,350)
\thicklines
\multiput(40,112.5)(15,0){16}{\line(1,0){10}}
\multiput(40,187.5)(15,0){16}{\line(1,0){10}}
\multiput(187.5,30)(0,15){16}{\line(0,1){10}}
\multiput(112.5,30)(0,15){16}{\line(0,1){10}}
\multiput(75,225)(150,0){2}{\circle*{10}}
\multiput(75,75)(150,0){2}{\circle*{10}}
\put(150,150){\circle*{10}}
\multiput(149,150)(1,0){3}{\line(1,1){75}}
\multiput(149,150)(1,0){3}{\line(1,-1){75}}
\multiput(74,75)(1,0){3}{\line(1,1){75}}
\multiput(74,225)(1,0){3}{\line(1,-1){75}}
\multiput(110,160)(75,-75){2}{\usebox{\sear}}
\put(110,110){\usebox{\near}}\put(160,160){\usebox{\swar}}
\end{picture}}

\newsavebox{\starss}\savebox{\starss}(350,350){
\begin{picture}(350,350)
\thicklines
\multiput(40,112.5)(15,0){16}{\line(1,0){10}}
\multiput(40,187.5)(15,0){16}{\line(1,0){10}}
\multiput(187.5,30)(0,15){16}{\line(0,1){10}}
\multiput(112.5,30)(0,15){16}{\line(0,1){10}}
\multiput(75,225)(150,0){2}{\circle*{10}}
\multiput(75,75)(150,0){2}{\circle*{10}}
\put(150,150){\circle*{10}}
\multiput(149,150)(1,0){3}{\line(1,1){75}}
\multiput(149,150)(1,0){3}{\line(1,-1){75}}
\multiput(74,75)(1,0){3}{\line(1,1){75}}
\multiput(74,225)(1,0){3}{\line(1,-1){75}}
\multiput(110,160)(75,-75){2}{\usebox{\sear}}
\put(185,185){\usebox{\near}}\put(85,85){\usebox{\swar}}
\end{picture}}

\phantom{a}

\vspace{2cm}
\centerline{\bf  Some exact results for the three-layer Zamolodchikov model.}

\vspace{1cm}

\phantom{a}
\centerline{ H.E. Boos\footnote{
on leave of absence from the Institute for
High Energy Physics, Protvino, 142284, Russia,
E-mail:
boos@th.physik.uni-bonn.de}}
\centerline{ Physikalisches Institut der Universit{\"a}t Bonn}
\centerline{ 53115, Bonn, Germany }

\phantom{a}

\vspace{0.5cm}

\phantom{a}

\centerline{ V.V. Mangazeev\footnote{E-mail:
vladimir@maths.anu.edu.au}}
\centerline{ Centre for Mathematics and its Applications,}
\centerline{  School of Mathematical Sciences,}
\centerline{ The Australian National University,}
\centerline{ Canberra, ACT 0200, Australia}

\vskip2em
\begin{abstract}
\noindent
In this paper we continue the study of the three-layer
Zamolodchikov model started in our previous works \cite{BM1} and
\cite{BM2}. We analyse numerically the solutions to the Bethe ansatz equations
obtained in \cite{BM2}. We consider two regimes I and II which differ
by the signs of the spherical sides $(a_1,a_2,a_3)
\rightarrow (-a_1,-a_2,-a_3)$. We accept the two-line hypothesis for
the regime I and the one-line hypothesis for the regime II. In the 
thermodynamic
limit we derive integral equations for distribution densities
and solve them exactly. We calculate the partition function for the
three-layer Zamolodchikov model and check a compatibility of this
result with the functional relations obtained in \cite{BM2}. We also
do some numeric checkings of our results.
\end{abstract}

\newpage

\section{Introduction}

This paper continues the investigation given in our previous papers
\cite{BM1}, \cite{BM2} of the three-layer Zamolodchikov
model \cite{Z1},\cite{Z2}. In paper \cite{Bax} Baxter considered
the Zamolodchikov model with some special boundary conditions.
Let us remind some basic ideas used by Baxter in \cite{Bax}.
To calculate the partition function Baxter used the symmetries
of the Boltzmann weights of the Zamolodchikov model and some its special property,
namely, the so-called "Body-Centered-Cube" (BCC) structure.
The last property allowed to write down the Boltzmann weights up to some gauge
and multiplication factors  in the following form
\be
\ov W(a|efg|bcd|h)=\sum_{\s}\phi_{\s,afch}e^{\s(K_1ag+K_2bf+K_3dh+K_4ce)}
\label{BCC}
\ee
with all spin variables $a\ldots h$ taking only two possible values $\pm 1$
and $\phi_{a,b}=-1$ if $a=b=-1$, $\phi_{a,b}=1$ otherwise.
The sum in (\ref{BCC}) is over the spin $\s=\pm 1$ placed in the center
of the cube corresponding to the weight $ \ov W(a|efg|bcd|h)$ as it is
shown in Fig.~1:

\addtolength{\unitlength}{-0.3\unitlength}

\begin{picture}(800,400)
\put(600,0){\bf Fig.1}
\put(450,30){
\begin{picture}(450,320)
\thicklines
\multiput(120,60)(0,200){2}{\line(1,0){200}}
\multiput(40,100)(40,0){5}{\line(1,0){20}}
\put(240,300){\line(0,-1){10}}
\put(40,300){\line(1,0){200}}
\multiput(120,60)(200,0){2}{\line(0,1){200}}
\put(40,100){\line(0,1){200}}
\multiput(240,100)(0,40){5}{\line(0,1){20}}
\multiput(120,60)(0,200){2}{\line(-2,1){80}}
\multiput(320,60)(-40,20){2}{\line(-2,1){30}}
\put(320,260){\line(-2,1){80}}
\put(120,60){\line(1,2){120}}
\qbezier(40,102)(180,182)(320,262)
\qbezier(40,100)(180,180)(320,260)
\qbezier(40,98)(180,178)(320,258)
\put(120,263){\line(3,-4){120}}
\put(120,260){\line(3,-4){120}}
\put(120,257){\line(3,-4){120}}
\qbezier(40,300)(180,180)(320,60)
\put(40,100){\circle*{12}}
\put(40,300){\circle*{12}}
\put(240,100){\circle*{12}}
\put(240,300){\circle*{12}}
\put(120,60){\circle*{12}}
\put(320,60){\circle*{12}}
\put(120,260){\circle*{12}}
\put(320,260){\circle*{12}}
\put(180,180){\circle*{15}}
\put(120,30){$e$}
\put(15,75){$c$}
\put(335,30){$d$}
\put(230,65){$h$}
\put(120,270){$a$}
\put(15,310){$g$}
\put(335,270){$f$}
\put(240,310){$b$}
\put(205,170){$\s$}

\put(260,140){\small $K_3$}
\put(280,210){\small $K_2$}
\put(80,160){\small $K_4$}
\put(80,220){\small $K_1$}
\end{picture}
}
\end{picture}

\vspace{1cm}

The coefficients $K_1,K_2,K_3$ and $K_4$ were written in terms of parameters
depending on the three angles $\theta_1,\theta_2,\theta_3$ and three spherical
sides $a_1,a_2,a_3$ of the spherical triangle:
\be
v_i=\tanh{2K_i}
\label{K}
\ee
where
\be
v_1=-z T_1 T_2,\quad v_2=-i z{{T_2}\over{T_1}},\quad
v_3=-z^{-1} T_1 T_2,\quad v_4=i z^{-1}{{T_2}\over{T_1}}
\label{v}
\ee
and
\be
z=\exp{(ia_3/2)}, \quad T_i = {[\tan{(\theta_i/2)}]}^{1/2}.
\label{zT}
\ee
The BCC-form allowed to write down the transfer matrix $T$ of the Zamolodchikov model as a
product of two matrices
\be
T_{\phi,\phi'}=\sum_{\phi''}X_{\phi,\phi''}\;Y_{\phi'',\phi'}
\label{TXY}
\ee
where
\be
X_{\phi,\phi''}=\prod_{\mbox{cubes}}\phi_{\s,ch}e^{\s(K_3dh+K_4ce)},
\quad
Y_{\phi'',\phi'}=\prod_{\mbox{cubes}}\phi_{\s,af}e^{\s(K_1ag+K_2bf)}
\label{XY}
\ee
and the summation in (\ref{TXY}) is over central spins $\phi''$ between two successive
layers of the lattice with the spins $\phi$ and $\phi'$ respectively.
Up to some simple multipliers defined by the formulae
(6.7-6.8) and (7.2-7.4) of \cite{Bax}
the Boltzmann weight $\ov W$ is equal to some other weight $W$
which is invariant under the action of the group $\Gamma$ of cube symmetries.
This group has two generating elements  $\tau$ and $\rho$ acting
on the set of spins $\{a|efg|bcd|h\}$ (see, also, \cite{KMS}) as follows
\be
\tau \{a|efg|bcd|h\} = \{a|feg|cbd|h\},
\quad
\rho \{a|efg|bcd|h\} = \{g|cab|fhe|d\}.
\label{rho}
\ee
Under these transformations the spherical angles  also change
\be
(\theta_1^{\tau},\theta_2^{\tau},\theta_3^{\tau})=(\theta_2,\theta_1,\theta_3),
\quad
(\theta_1^{\rho},\theta_2^{\rho},\theta_3^{\rho})=(\pi-\theta_1,\theta_3,\pi-\theta_2).
\label{rhoang}
\ee
Let us go back to the BCC-form of the weight (\ref{BCC}) and the expression
(\ref{TXY}) for the transfer matrix. Unfortunately,
two matrices $X$ and $Y$ do not commute with each other.
In order to handle this difficulty  Baxter
modified the boundary conditions for the
central spins of Boltzmann weights so that each central spin $\s$ was replaced
by the product of a pair of spins $\mu\mu'$
as it is shown in Fig.~2

\begin{picture}(800,360)
\put(200,0){
\begin{picture}(350,320)
\thicklines
\multiput(120,60)(0,200){2}{\line(1,0){200}}
\multiput(40,100)(40,0){5}{\line(1,0){20}}
\put(240,300){\line(0,-1){10}}
\put(40,300){\line(1,0){200}}
\multiput(120,60)(200,0){2}{\line(0,1){200}}
\put(40,100){\line(0,1){200}}
\multiput(240,100)(0,40){5}{\line(0,1){20}}
\multiput(120,60)(0,200){2}{\line(-2,1){80}}
\multiput(320,60)(-40,20){2}{\line(-2,1){30}}
\put(320,260){\line(-2,1){80}}
\put(120,60){\line(1,2){120}}
\qbezier(40,102)(180,182)(320,262)
\qbezier(40,100)(180,180)(320,260)
\qbezier(40,98)(180,178)(320,258)
\put(120,263){\line(3,-4){120}}
\put(120,260){\line(3,-4){120}}
\put(120,257){\line(3,-4){120}}
\qbezier(40,300)(180,180)(320,60)
\put(40,100){\circle*{12}}
\put(40,300){\circle*{12}}
\put(240,100){\circle*{12}}
\put(240,300){\circle*{12}}
\put(120,60){\circle*{12}}
\put(320,60){\circle*{12}}
\put(120,260){\circle*{12}}
\put(320,260){\circle*{12}}
\put(180,180){\circle*{15}}
\put(120,30){$e$}
\put(15,75){$c$}
\put(335,30){$d$}
\put(230,65){$h$}
\put(120,270){$a$}
\put(15,310){$g$}
\put(335,270){$f$}
\put(240,310){$b$}
\put(205,170){$\s$}

\end{picture}
}
\put(700,0){
\begin{picture}(300,320)
\thicklines
\multiput(120,60)(0,200){2}{\line(1,0){200}}
\multiput(40,100)(40,0){5}{\line(1,0){20}}
\put(240,300){\line(0,-1){10}}
\put(40,300){\line(1,0){200}}
\multiput(120,60)(200,0){2}{\line(0,1){200}}
\put(40,100){\line(0,1){200}}
\multiput(240,100)(0,40){5}{\line(0,1){20}}
\multiput(120,60)(0,200){2}{\line(-2,1){80}}
\multiput(320,60)(-40,20){2}{\line(-2,1){30}}
\put(320,260){\line(-2,1){80}}

\multiput(120,60)(-80,40){2}{\line(1,1){200}}
\multiput(320,60)(-80,40){2}{\line(-1,1){200}}
\multiput(40,100)(30,10){6}{\line(3,1){20}}
\multiput(140,200)(30,10){6}{\line(3,1){20}}
\multiput(220,160)(-40,20){2}{\line(-2,1){30}}
\multiput(220,160)(10,-30){2}{\line(1,-3){7}}
\multiput(120,260)(10,-30){2}{\line(1,-3){7}}

\put(220,160){\circle*{10}}
\put(140,200){\circle*{10}}

\put(40,100){\circle*{12}}
\put(40,300){\circle*{12}}
\put(240,100){\circle*{12}}
\put(240,300){\circle*{12}}
\put(120,60){\circle*{12}}
\put(320,60){\circle*{12}}
\put(120,260){\circle*{12}}
\put(320,260){\circle*{12}}
\put(120,30){$e$}
\put(15,75){$c$}
\put(335,30){$d$}
\put(230,65){$h$}
\put(120,270){$a$}
\put(15,310){$g$}
\put(335,270){$f$}
\put(240,310){$b$}

\put(245,155){\small $\mu$}
\put(90,190){\small $\mu'$}


\end{picture}
}
\multiput(600,180)(0,1){3}{\line(1,0){100}}
\put(660,160){\usebox{\eara}}
\put(630,0){\bf Fig.2}
\end{picture}

$\quad$\\
The matrices $X$ and $Y$ become
\be
X_{\phi,\phi''}=\prod_{\mbox{cubes}}\phi_{\mu\mu',ch}e^{\mu\mu'(K_3dh+K_4ce)},
\quad
Y_{\phi'',\phi'}=\prod_{\mbox{cubes}}\phi_{\mu\mu',af}e^{\mu\mu'(K_1ag+K_2bf)}.
\label{XYmod}
\ee
Now there exist  some non-singular matrices $P,Q,R,S$ such that
\be
X(v_3,v_4)=P(\th_1)R(v_3,v_4)Q^{-1}(\th_1),\quad Y(v_1,v_2)=Q(\th_1)S(v_1,v_2)P^{-1}(\th_1)
\label{RS}
\ee
$\quad$\\
and the matrices $R(v_3,v_4)$ and $S(v_1,v_2)$ are diagonal.
As was mentioned above it leads to a modification of boundary conditions for the initial
spins $\s$, namely, the product of all $\s$'s along the front-to-back direction should be 1:
\be
\s\s'\s''...=\mu\mu'\mu'\mu''\mu''\mu'''...=1,
\label{spinprod}
\ee
while the initial $\s$'s were not constrained at all.
If the cubic lattice is infinite in all three directions
one can assume  that the modification of the boundary conditions
does not change the partition function. Further Baxter used two symmetries
which do not affect front-to-back direction of the lattice.
The first one is just the transformation $\tau$ from (\ref{rho}).
The second one, say $\tau'$, is the reflection interchanging
left and right
\be
\tau' \{a|efg|bcd|h\} = \{f|dab|ghe|c\}
\label{tau'}
\ee
with
\be
(\theta_1^{\tau'},\theta_2^{\tau'},\theta_3^{\tau'})=
(\pi-\theta_1,\theta_2,\pi-\theta_3).
\label{tau'ang}
\ee
Using these symmetries
Baxter derived two functional relations for the largest diagonal
elements of the matrices $R$ and $S$ in (\ref{RS})
and found its solution. It allowed him to come to the
marvelous result for the partition function per site $\kappa$
(formula (8.14) and (9.7) of \cite{Bax})
\footnote{Here we have omitted
the normalization function $\xi$ defined by formula (7.2) of \cite{Bax}}:
\be
\ln{\kappa/2}=J(\zeta_1)+J(\zeta_2)+J(\zeta_3)+J(\zeta_4)
\label{kappa}
\ee
where
\be
J(\zeta)=(2\pi)^{-1}\int_{0}^{\zeta}(\ln{(2\cos{x})}+x\tan{x})dx
\label{J}
\ee
and
\be
e^{i\zeta_1}=-i/v_1,\quad e^{i\zeta_2}=i v_2,\quad
e^{i\zeta_3}=-iv_3,\quad e^{i\zeta_4}=i/v_4.
\label{zeta}
\ee
In paper \cite{Bax0} Baxter obtained the manifest result
for the isotropic case $\theta_i=a_i=\pi/2$
\be
\kappa=2^{3/4}\,(2^{1/2}-1)\,e^{2\,G/\pi}=1.2480...
\label{kappaisot}
\ee
where $G=0.915965...$ is Catalan constant.
In papers \cite{BF1},\cite{BF2}
Baxter and Forrester checked this result numerically using the variational method.

In \cite{Bax} Baxter also obtained a generalization of the
result (\ref{kappa}-\ref{J}) to the case of the lattice which is infinite
in the vertical and left-to-right directions and has $n$
layers in the front-to-back direction. The claim is that the formula (\ref{kappa})
is still correct but the function $J$ given by (\ref{J}) has to be replaced by
the function $J_n$ for which Baxter found the following formula:

\be
J_n(\zeta)=\hat J_n(\zeta)+{1\over 4}
\ln{
\tan{
({{\zeta}\over{2}}+{{\pi}\over{4}})
}
}
\label{J_n}
\ee
with the function $\hat J_n$
\be
\hat J_n(\zeta)=\int_{-\infty}^{\infty}
{
{\sinh{2x\zeta}}\over{4x\sinh{\pi x}}
}
(\mbox{cosech}(2\pi x)-n^{-1}(\mbox{cosech}(2\pi x/n))dx.
\label{J_nhat}
\ee
Baxter also checked that this result was valid for three particular cases, namely,
$n=1$ when the model was trivial, $n=2$ when the model became the planar free-fermion
model and for $n=\infty$ i.e. the solution (\ref{J_n}-\ref{J_nhat}) reduced to the
formula (\ref{J}) in the limit $n\rightarrow\infty$.

In this paper we would like to examine this result for the case $n=3$ using the Bethe
ansatz technique developed in our previous works
\cite{BM1}, \cite{BM2}. The formula (\ref{J_n}) turns out to be incorrect for $n=3$.
We think that the main reason is that in the case when $n=3$ one can not apply
both of the symmetries $\tau$ and $\tau'$ to derive the functional relations and
the resulting formula (\ref{kappa}). As we show in section 2 the symmetry $\tau'$
fails for $n=3$. We expect that it also happens for any odd $n>1$. Other reason is
that the function $J_n$ should have more complicated analytic properties in comparison
with that assumed by Baxter, namely, that $J_n(\zeta)$ is analytic in the strip
$|\mbox{Re}(\zeta)|<\pi/2$. In section 9 we discuss the analytic properties of our solution
and check the validity of the inversion relation.

One should note that the case of even $n$ seems to be different.
We have checked numerically that
for $n=4,6$ both of the symmetries $\tau$ and $\tau'$ work. So we expect
that they also work for any even $n$ and
the symmetry method by Baxter gives
the right answer.

In paper \cite{BB} Baxter and Bazhanov found a generalization of the
Zamolodchikov model to the case of the arbitrary number
of states $N$ and obtained the equivalence of the $n$-layer model with modified
boundary conditions  and the $sl(n)$-chiral
Potts model at $q^{2N}=1$, \cite{BKMS},\cite{DJMM}.
In particular, the modified three-layer Zamolodchikov model appeared
to be equivalent to the $sl(3)$-chiral Potts model at $q^2=-1$.

The main purpose of this paper is to calculate the partition function of the
modified three-layer Zamolodchikov model or the $sl(3)$-chiral Potts model at $q^2=-1$
in the thermodynamic limit using
the Bethe ansatz technique.
We also perform extensive numerical calculations to support our results.

The paper is organized as follows. In section 2 we discuss the symmetry properties
of the modified three-layer Zamolodchikov model with respect to the
$\tau$ and $\tau'$ transformations.
In section 3 we establish an exact correspondence between the
three-layer Zamolodchikov model and the homogeneous $sl(3)$-chiral Potts model.
In section 4 we discuss the Bethe ansatz equations
corresponding to the largest eigenvalues of the transfer matrices.
In section 5 we consider some special properties of the transfer matrix spectrum.
In section 6 we introduce two-line hypothesis of solutions to the
Bethe ansatz equations. We also derive and solve the integral equations
for the distribution densities in the thermodynamic limit.
In section 7 we calculate the partition function in thermodynamic limit
using the solution for the distribution densities.
A compatibility of the results from section 6 with the functional relations
is established in section 8.  In section 9 we consider the mechanism how the
inversion relations for the partition function work.
In section 10 we consider one-line regime.
In the last section we discuss the results and outline some further developments.

\section{The $\t$ and $\t'$ transformations for the three-layer
Boltzmann weight}

As was explained in \cite{BB} the Boltzmann weights for the $n$-layer
Baxter-Bazhanov model with $N$ states and the $sl(n)$-chiral Potts model at $q^{2N}=1$
can be connected with each other
(see formulae (2.20) and (2.21) of \cite{BB}). In particular,
for the three-layer Zamolodchikov model we have
\be
W_{zam}(\{\a\},\{\b\},\{\g\},\{\d\})
=\prod_{i=1}^3\ov W(\d_i|\a_i,\g_i,\d_{i+1}|\g_{i+1},\a_{i+1},\b_i|\b_{i+1})
\label{Wzam0}
\ee
where $\{\a\}=\{\a_1,\a_2,\a_3\}, \{\b\}=\{\b_1,\b_2,\b_3\},\ldots$
Here we imply periodic boundary conditions in
the front-to-back direction and each $\ov W^i$ is given by (\ref{BCC}).
The Boltzmann weight (\ref{Wzam0}) can be
decomposed as follows
\be
W_{zam} = W_+ + W_-
\label{Wzam}
\ee
where $W_+$  corresponds to the modified model described above with the constraint
$$
\s\s'\s''=1
$$
on the central spins of the weights $\ov W^i$
and the weight $W_-$ corresponds to a different choice of sign
$$
\s\s'\s''=-1.
$$
Using formulas from the previous section we can write
$$
W_{\pm}(\{\a\},\{\b\},\{\g\},\{\d\})=
$$
$$
\sum_{\{\mu\}}
\phi_{\a_1,\pm\mu_1\mu_3}\phi_{\a_2,\mu_2\mu_1}\phi_{\a_3,\mu_3\mu_2}
e^{K_4(\a_1\a_2\mu_1\mu_2\;+\;\a_2\a_3\mu_2\mu_3\;\pm\;\a_3\a_1\mu_3\mu_1)}
$$
$$
\phi_{\b_1,\pm\mu_1\mu_3}\phi_{\b_2,\mu_2\mu_1}\phi_{\b_3,\mu_3\mu_2}
e^{K_3(\b_1\b_2\mu_1\mu_2\;+\;\b_2\b_3\mu_2\mu_3\;\pm\;\b_3\b_1\mu_3\mu_1)}
$$
$$
\phi_{\g_1,\mu_1\mu_2}\phi_{\g_2,\mu_2\mu_3}\phi_{\g_3,\pm\mu_3\mu_1}
e^{K_2(\g_1\g_2\mu_1\mu_2\;+\;\g_2\g_3\mu_2\mu_3\;\pm\;\g_3\g_1\mu_3\mu_1)}
$$
\be
\phi_{\d_1,\mu_1\mu_2}\phi_{\d_2,\mu_2\mu_3}\phi_{\d_3,\pm\mu_3\mu_1}
e^{K_1(\d_1\d_2\mu_1\mu_2\;+\;\d_2\d_3\mu_2\mu_3\;\pm\;\d_3\d_1\mu_3\mu_1)}.
\label{Wpm}
\ee
Now let us consider the $\tau$-transformation from (\ref{rho}). Under this
transformation $\{\a\}\leftrightarrow\{\g\}$ and $\theta_1\leftrightarrow\theta_2$ or
$K_1,K_2,K_3,K_4\rightarrow K_1,K_4-i\pi/4,K_3,K_2+i\pi/4$.
Using the following relation
$$
\phi_{\a_1,\mu_1\mu_2}\phi_{\a_2,\mu_2\mu_3}\phi_{\a_3,\e\,\mu_3\mu_1}
e^{-i{\pi\over4}(\a_1\a_2\mu_1\mu_2\;+\;\a_2\a_3\mu_2\mu_3\;+\e\;\a_3\a_1\mu_3\mu_1)}=
$$
\be
\phi_{\a_2,\mu_1\mu_2}\phi_{\a_3,\mu_2\mu_3}\phi_{\a_1,\e\,\mu_3\mu_1}
e^{i{\pi\over4}(\a_1\a_2\;+\;\a_2\a_3\;+\e\;\a_3\a_1\;-\;
\mu_1\mu_2\;-\;\mu_2\mu_3\;-\;(2-\e)\,\mu_3\mu_1\;-\;(1+2\e))}
\label{eq}
\ee
with $\e=\pm 1$
which can be directly verified we get the $\tau$-transformed weights $W_{\pm}$:
\be
W_{\pm}^{\tau}(\{\a^{\tau}\},\{\b^{\tau}\},\{\g^{\tau}\},\{\d^{\tau}\})=
{{{\it F}(\{\a\})}\over{{\it F}(\{\g\})}}W_{\pm}(\{\a\},\{\b\},\{\g\},\{\d\})
\label{Wtau}
\ee
where $\{\a^{\tau}\},\{\b^{\tau}\},\{\g^{\tau}\},\{\d^{\tau}\}=
\{\g\},\{\b\},\{\a\},\{\d\}$ and
$$
{\it F}(\{\a\})=e^{i{\pi\over4}(\a_1\a_2+\a_2\a_3+\a_3\a_1)}.
$$
From (\ref{Wtau}) we see that  $W_{+}$ and $W_{-}$ transform in
the same way. The ratio of the functions ${\it F}$ is the gauge
transformation, i.e. up to the same gauge transformations both weights
$W_{+}$ as well as $W_{-}$ are $\tau$-invariant.
Of course, their sum $W_{zam}=W_{+}+W_{-}$
is also $\tau$-invariant.

As for the second transformation $\tau'$ given by (\ref{tau'}) the sum $W_{+}+W_{-}$
is again invariant (up to some gauge)
under $\tau'$. However, each term $W_{\pm}$ is not invariant under $\tau'$ separately.
It means that for $n=3$ only the first $\tau$-symmetry is valid for the modified model
but not $\tau'$. We have checked numerically that the same situation takes place for $n=5$.
We can expect that for any odd $n>1$ the symmetry $\tau'$ is not
valid. Therefore, for odd values of $n>1$ one can not use both symmetries
$\tau$ and $\tau'$ for a calculation  of  the partition function.
However, as was mentioned in the Introduction both of the symmetries $\tau$ and $\tau'$
seem to work for any even $n$.

\section{The equivalence to the $sl(3)$-chiral Potts model}

As was shown in \cite{BB} the modified model with the Boltzmann weights $W_{+}$
corresponds to the $sl(3)$-chiral Potts model at $q^2=-1$. In
\cite{BM1},\cite{BM2} we used the expression for the "star" weight defined in
\cite{KMN}. It differs from the definition of \cite{BB} (see the formula (3.14) there)
by a rotation of the star weight over the angle ${\pi}\over 2$ round the front-to-back
direction. However, for the modified model with $n=3$
the weight $W_+$ is not invariant under this rotation (as in the case of $\tau'$).

Therefore, we need a direct connection of the Boltzmann weight for the modified
model $W_{+}$ given by (\ref{Wpm}) with the definition (2.17) from \cite{BM2}
\be
W^{star}(\a,\b,\g,\d;p,p',q,q')\,=\,\sum_{\mu}
{
{{\ov W_{pq}}(\a,\mu)\,{\ov W_{p'q'}}(\g,\mu)\,
{\ov W_{q'p}}(\mu,\b)}
\over
{{\ov W_{p'q}}(\d,\mu)}
},
\label{star0}
\ee
where the rapidity picture is defined by Fig.~3:

\begin{picture}(700,350)
\put(250,70)
{\begin{picture}(300,150)
\thicklines
\multiput(0,0)(40,40){4}{\line(1,1){30}}
\multiput(150,0)(-40,40){4}{\line(-1,1){30}}
\put(120,120){\usebox{\near}}\put(-12,120){\usebox{\nwar}}
\multiput(75,0)(0,150){2}{\circle*{10}}
\multiput(74,0)(1,0){3}{\line(0,1){150}}
\multiput(74,120)(1,0){3}{\line(-1,-2){10}}
\multiput(74,120)(1,0){3}{\line(1,-2){10}}
\put(-20,30){ $p$}\put(140,30){ $q$}
\put(80,-10){ $\a$}\put(80,160){ $\b$}
\put(150,75){\bf {$=\quad \overline W_{pq}(\a,\b)$}}
\end{picture}}
\put(600,-10){\bf Fig. 3}
\put(750,70)
{\begin{picture}(300,150)
\thicklines
\multiput(0,150)(40,-40){4}{\line(1,-1){30}}
\multiput(150,150)(-40,-40){4}{\line(-1,-1){30}}
\put(0,0){\usebox{\swar}}\put(108,0){\usebox{\sear}}
\multiput(75,0)(0,150){2}{\circle*{10}}
\multiput(74,0)(1,0){3}{\line(0,1){150}}
\multiput(74,120)(1,0){3}{\line(-1,-2){10}}
\multiput(74,120)(1,0){3}{\line(1,-2){10}}
\put(-20,30){ $p$}\put(140,30){ $q$}
\put(80,-10){ $\a$}\put(80,160){ $\b$}
\put(150,75){{$=\quad{\displaystyle 1 \over
{\displaystyle\phantom{I}{ \overline W_{qp}(\a,\b)}^{\phantom{I}}}}$}}
\end{picture}}
\end{picture}

\vspace{1cm}
In (\ref{star0}) all indices  are twofold, for example,
$\a=(\a_1,\a_2)$ where $\a_1,\a_2=\{0,1\}$.
Let us denote these indices as $\tilde\a_i,
\tilde\b_i$ etc.
A correspondence with indices taking values $\pm1$ is very simple:
$\tilde\a_i=h(\a_i)$, $ h(\a)={{1-\a}\over 2}  $.

A direct calculation shows that
\beq
&{\ds W_{+}(\{\a\},\{\b\},\{\g\},\{\d\})=
2\,e^{3(K_1+K_2+K_3+K_4)}{{(-1)^{h(\d_1\d_2)+h(\d_2\d_3)+h(\d_1\d_2)h(\d_2\d_3)}}
\over{(-1)^{h(\g_1\g_2)+h(\g_2\g_3)+h(\g_1\g_2)h(\g_2\g_3)}}}\times}& \nonumber\\
&{\ds W^{star}(\{h(\a_1\a_2),h(\a_2\a_3)\},\{h(\b_1\b_2),h(\b_2\b_3)\}
,\{h(\g_1\g_2),h(\g_2\g_3)\},\{h(\d_1\d_2),h(\d_2\d_3)\};p,p',q,q')}& \nonumber\\
\label{W+Wstar}
\eeq
where the coefficient $2$ is due to the fact that in (\ref{star0})
a summation is over the twofold index
while  in (\ref{Wpm}) indices are threefold.
In addition, the parameters $p,p',q,q'$ should be so that
$$
e^{4\,K_1}={{1+v_1}\over{1-v_1}}=-{{1-p'/q}\over{1+p'/q}},\quad
e^{4\,K_2}={{1+v_2}\over{1-v_2}}=-{{1+p'/q'}\over{1-p'/q'}},
$$
\be
e^{4\,K_3}={{1+v_3}\over{1-v_3}}=-{{1+q'/p}\over{1-q'/p}},\quad
e^{4\,K_4}=\;{{1+v_4}\over{1-v_4}}\;=\;\;{{1+p/q}\over{1-p/q}}
\label{Kpqp'q'}
\ee
and
\be
v_1=-{q\over{p'}},\quad v_2={q'\over{p'}},\quad v_3={p\over{q'}},\quad
v_4={p\over{q}}
\label{vpq}
\ee
with the parameters $v_i$ given by the formula (\ref{v}).
Without loss of generality we can choose
\be
q=1, \quad q'=-i/T_1^2,\quad p=v_4,\quad p'=-{1\over{v_1}}
\label{qq'}
\ee
in accordance with formulae (\ref{vpq}) and with $T_1$ defined in (\ref{zT}).

\section{The partition function and Bethe ansatz equations}

Our main goal is to calculate the partition function $Z$ for the modified Zamolodchikov
model with three layers
at least in the thermodynamic limit. If the lattice is finite and
has $N$ sites in the left-to-right direction and $M$ sites in the vertical
direction (in the front-to-back direction we have three sites) then
\be
Z=\sum_I {t_+}(K_1,K_2,K_3,K_4)_{I}^M
\label{Sigma}
\ee
where ${t_+}(K_1,K_2,K_3,K_4)_{I}$ is the $I$-th eigenvalue of the $N$-site transfer
matrix for the modified Zamolodchikov model with the weights $W_+$ given by
(\ref{Wpm}) and the sum is taken over all eigenvalues.
If the number $M$ tends to infinity only the eigenvalue with maximal absolute
value contributes
(see \cite{BaxB}) and for the logarithm of the partition function per site
$\kappa=Z^{1/(3NM)}$
we have
\be
\ln{\kappa}={1\over{3N}}\ln{{t_+}(K_1,K_2,K_3,K_4)_{0}}.
\label{kappa1}
\ee
Now we have to connect the eigenvalues ${t_+}(K_1,K_2,K_3,K_4)$ with
the eigenvalues $t(p;q,q')$, $\ov t(p';q,q')$ of the
$N$-site transfer matrices $T(p;q,q')$
and $\ov T(p';q,q')$ of the $sl(3)$-chiral Potts model which
were written in our previous work \cite{BM2} as follows
\be
t(p;q,q')\,=\,{{2^N}\over{(p\,+\,q)^N\;(p\,+\,q')^N}}\,s(p;q,q'),\quad
{\ov t(p;q,q')}\,=\,{{2^N}\over{(p\,-\,q)^N\;(p\,+\,q')^N}}\,
{\ov s(p;q,q')}
\label{tt}
\ee
where the functions $s(p;q,q')$ and $\ov s(p;q,q')$ are  polynomials in $p$ of
the degree $k$
\be
s(p;q,q')\,=\,a_k(q,q')\,\prod_{{\it i}=1}^k\,(p\,-\,p_{\it i}),\quad
{\ov s(p;q,q')}\,=\,\ov a_k(q,q')\,\prod_{i=1}^k\,(p\,-\,{\ov p_i})
\label{sts}
\ee
and $p_i,\ov p_i$ are two sets of $k$ numbers depending only on $q,q'$.
Due to the correspondence (\ref{W+Wstar}) we have
\be
t_+(K_1,K_2,K_3,K_4)=2^N e^{3N(K_1+K_2+K_3+K_4)}t(p;q,q')\ov t(p';q,q').
\label{t+}
\ee
Hence,
\be
\ln{(\kappa/2)}=-{{2\ln{2}}\over3}+
K_1+K_2+K_3+K_4+{1\over{3N}}\ln{t(p;q,q')_0\;\ov t(p';q,q')_0}.
\label{kappa2}
\ee
where $t(p;q,q')_0$ and $\ov t(p';q,q')_0$ are the eigenvalues of the
transfer matrices $T(p;q,q')$ and $\ov T(p';q,q')$ which give
the maximal absolute value of the product
$t(p;q,q')_0 \ov t(p';q,q')_0$.
As was discussed in \cite{BM2} the degree of the polynomials $k$ can be only
$2N$ or $2N-1$.
In fact, we need to know only the product of unknown functions
$a_n(q,q'),\ov a_n(q,q')$ which enter in (\ref{sts}), namely,
\be
a_{2N}(q,q')\,\ov a_{2N}(q,q')=4, \quad
a_{2N-1}(q,q')\,\ov a_{2N-1}(q,q')=N\,(q'^2-q^2)
\label{a2N}
\ee
to determine uniquely the eigenvalue $t_+(K_1,K_2,K_3,K_4)$.
The parameters $p_i$ and $\ov p_i$ should satisfy the Bethe ansatz equations
(5.4-5.5) of
\cite{BM2}
\be
{{f(p_i,\om^{\pm1},-q)}^N\over {f(p_i,\om^{\pm1},-q')}^N}\,=\,
(-1)^{k-1}
\prod_{j=1}^k{p_i+\om^{\mp1}\ov p_j\over p_i-\om^{\mp1}\ov p_j}
\label{bethe1}
\ee
and
\be
{{f(\ov p_i,\om^{\pm1},\;q)}^N\over {f(\ov p_i,\om^{\pm1},-q')}^N}=
(-1)^{k-1}
\prod_{j=1}^k{\ov p_i+\om^{\mp1}p_j\over \ov p_i-\om^{\mp1}p_j}
\label{bethe2}
\ee
where
\be
f(p,x,q)\,=\,{{p\,-\,x\,q}\over{p\,+\,q}}.
\label{f}
\ee
and $\om=e^{2i\pi/3}$ is the root of unity of power three.

Below we shall consider two regimes I and II. Let us remind
that the Boltzmann weights depend on spherical angles $(\th_1,\th_2,\th_3)$
and spherical sides $(a_1,a_2,a_3)$ via the parameters $v_i$
given by the formulae (\ref{v}) and (\ref{zT}).
Let us say that this parameterization corresponds to the regime I.

The second regime II will correspond to  the negating of spherical sides i.e.
$(a_1,a_2,a_3)\rightarrow (-a_1,-a_2,-a_3)$ with the spherical angles
$(\th_1,\th_2,\th_3)$ being unchanged. In formulae for $v_i$ only one substitution
should be taken $z\rightarrow z^{-1}$ i.e. new $v'_i$ are given by
\be
v'_1=-z^{-1}\,T_1\,T_2,\quad v'_2=-i z^{-1}\,T_2/T_1,\quad
v'_3=-z\,T_1\,T_2,\quad v'_4=i z\,T_2/T_1.
\label{vv}
\ee
However,  we do not have a good geometric picture  in regime II.

\section{The structure of the spectrum of the transfer matrix}

Let us introduce some notations we use below.   There are three
mutually commutative operators
$X_1,X_2,X_3$ which also commute with both transfer matrices
$T$ and $\ov T$
\be
[X_i,T(p;q,q')]\;=\;[X_i,\ov T(p;q,q')]\;=\;0\quad \mbox{with}\quad
X_i\;=\;x_i\otimes x_i\otimes\ldots \otimes x_i
\label{XTT}
\ee
where
\be
x_1\;=\; \left(\matrix{
0&0&1&0\cr
0&0&0&1\cr
1&0&0&0\cr
0&1&0&0\cr}\right)\quad
x_2\;=\; \left(\matrix{
0&1&0&0\cr
1&0&0&0\cr
0&0&0&1\cr
0&0&1&0\cr}\right)\quad
x_3\;=\; \left(\matrix{
0&0&0&1\cr
0&0&1&0\cr
0&1&0&0\cr
1&0&0&0\cr}\right).
\label{x123}
\ee

In addition, the transfer matrices $T$ and $\ov T$ commute with
shift operators $P_2$ and $P_3$ which act in left-to-right and
front-to-back directions respectively. However, only the operator
\be
\ov X=X_1+X_2+X_3
\ee
commutes with both $P_2$ and $P_3$. As a result all eigenvectors split
into two big classes $\psi^+$ and $\psi^-$
\be
X_i\psi^+=\psi^+,\quad i=1,2,3,\quad \ov X\psi^+=3\psi^+,
\quad \ov X\psi^-=-\psi^-
\ee
\be
P_2\psi^\pm=\Omega^{p_2}\psi^\pm, \quad \Omega^N=1, \quad p_2=0,...,N-
\ee
\be
P_3\psi^\pm=\om^{p_3}\psi^\pm,\quad \om^3=1, \quad p_3=-1,0,1
\ee

It is interesting to note that all eigenvalues corresponding to
the vectors $\psi^+$ are in the sector with the degree $k=2N$ of
the polynomials $s(p)$ and $\ov s(p)$ and non-degenerate in general
case. In contrary, the eigenvalues corresponding to $\psi^-$
are in the sector with $k=2N-1$ and have multiplicity 3 in accordance
with three eigenvalues of $P_3$, $p_3=-1,0,1$.
So let us use the following signature for eigenstates with
fixed momentums $p_2$ and $p_3$:
$(p_2;p_3)^+$ for $\psi^+$ and $(p_2;p_3)^-$ for $\psi^-$.

Below we give  a complete spectrum of the transfer matrix
$T(p;q,q')\ov T(p';q,q')$ for the cases $N=2,3,4$ and largest
eigenvalues in sectors $(0;0)^+$ and $(0;*)^-$ (hereafter $*$ stands
for the values of $p_3=-1,0,1$)
for the cases $N=5,6,7,8,9$  with a special
choice of the spectral parameters
\be
q=1,\quad q'=-i,\quad p=e^{i\pi\over4},\quad p'=e^{-i\pi/4}.
\ee
It simplifies drastically the structure of the spectrum
and corresponds to the symmetric point due to (\ref{qq'})
\be
\th_1={\pi\over2},\quad\th_2={\pi\over2},\quad\th_3={\pi\over2}.
\ee
Numerical calculations show that such a choice  does not change
a multiplicity of the largest eigenvalue but
it leads to some additional degeneracy in low eigenvalues. In general case
the number of different eigenvalues grows too rapidly with $N$.

In tables 1,2,3 we present the absolute eigenvalues $\Lambda$
\be
\Lambda= |t(p;q,q')\ov t(p';q,q')|
\ee
of the transfer matrix $T(p;q,q')\ov T(p';q,q')$ for the cases $N=2,3,4$ respectively.
The upper subscript of eigenvalues shows its multiplicity.
\vspace{0.3cm}

\begin{center}
\begin{tabular}{|c|c|c|}
\hline $(p_2;p_3)^\e$ & $\Lambda_1$ & $\Lambda_2$ \\
\hline $(0;0)^+$  & $16^{2}$& $-$ \\
\hline $(0;\pm1)^+$& $80$& $-$ \\
\hline $(0;*)^-$ & $16(3+2\sqrt{2})\approx{\bf 93.2548}$ &
$16(3-2\sqrt{2})\approx2.7452$ \\
\hline $(1;*)^-$  & $16^{2}$& $-$ \\
\hline
\end{tabular}
\end{center}
\begin{center}
Table 1. The case $N=2$.
\end{center}
\vspace{0.2cm}

\begin{center}
\begin{tabular}{|c|c|c|c|c|}
\hline $(p_2;p_3)^\e$& $\Lambda_1$ & $\Lambda_2$ & $\Lambda_3$
& $\Lambda_4$ \\
\hline $(0;0)^+$ & {\bf 2137.2}& 38.81 & $32^{2}$& $-$\\
\hline $(0;\pm1)^+$ & 96& $-$& $-$ & $-$\\
\hline $(1;0)^+,(2;0)^+$ & 96& $-$& $-$ & $-$\\
\hline $(1;\pm1)^+,(2;\pm1)^+$ & 317.62& 250.91 & $-$& $-$ \\
\hline $(0;*)^-$&$288^2$ & 119.43& $96^{2}$ & 8.574 \\
\hline $(1;*)^-,(2;*)^-$&$1120$ & $221.267^{2}$&
$13.884^{2}$
& $-$\\
\hline
\end{tabular}
\end{center}
\begin{center}
Table 2. The case $N=3$.
\end{center}
\vspace{0.2cm}

\begin{center}
\begin{tabular}{|c|c|c|c|c|c|c|c|c|}
\hline $(p_2;p_3)^\e$& $\Lambda_1$ & $\Lambda_2$ & $\Lambda_3$
& $\Lambda_4$& $\Lambda_5$& $\Lambda_6$& $\Lambda_7$& $\Lambda_8$ \\
\hline $(0;0)^+$  & $5326.8^{2}$& 1840.5 & 1381.9 & 256&
105.5& $49.21^{2}$& $-$& $-$\\
\hline $(0;\pm1)^+$ & 4920.0& $2720.5^{2}$ &
$1456.2$ & $852.5^{2}$& $135.8$& $-$& $-$& $-$\\
\hline $(1;0)^+$,$(3;0)^+$ & $3002.8^{2}$& $164.77^{2}$
& $-$ & $-$& $-$& $-$& $-$& $-$\\
\hline $(2;0)^+$ & $9741.8^{2}$& $256^{2}$
& $242.2^{2}$ & $-$& $-$& $-$& $-$& $-$\\
\hline $(1;\pm1)^+,(3;\pm1)^+$ & 1858.2& 1313.8& $522.86$&
$229.6$ & $-$& $-$& $-$& $-$\\
\hline $(2;\pm1)^+$ & 6269.5& $627.07^4$& $381.53$&
$-$ & $-$& $-$& $-$& $-$\\
\hline $(0;*)^-$&${\bf 13548.7}$ & $4518.5^2$& $2304^2$ &
$1062.1^2$&$1024^3$& $256^3$ & $127.6^2$& $19.4$\\
\hline $(1;*)^-,(3;*)^-$&$7952.6^2$ & $3694.9^{2}$&
$1386.3^{2}$
& $603.3^2$& $582.7^2$& $467.1^2$& $287.1^2$& $28.2^2$\\
\hline $(2;*)^-$&$2458.9^4$ & $1536^{2}$&
$1511.2^{2}$
& $338.2^4$& $271.9^2$& $40.8^2$& $-$& $-$\\
\hline
\end{tabular}
\end{center}
\begin{center}
Table 3. The case $N=4$.
\end{center}
In spite of the fact that there are complex eigenvalues in the spectrum,
all eigenvalues  in the sector $(0;0)^+$ are real and positive.
Bold numbers correspond to the eigenvalues with largest absolute value.

In table 4 we give dimensions $m$ and numeric values
$
\l={1\over3N}\log\Lambda_1
$
 only for the sectors $(0;0)^+$ and $(0;*)^-$ with $N=2,\ldots,9$.

\begin{center}
\begin{tabular}{|c|c|c|c|c|c|c|c|c|c|}
\hline
\hline $(p_2;p_3)^\e$& & $N=2$ & $N=3$ & $N=4$
& $N=5$& $N=6$& $N=7$& $N=8$& $N=9$ \\
\hline $(0;0)^+$ & $m$ & $2$& 4 & 8 & 18&
66& $196$& $696$& $2440$\\
\cline{2-10} & $\l$ & $0.4621$ & ${\bf0.8519}$ &
$0.7150$ & $0.7345$& ${\bf0.8263}$& $0.7771$& $0.7827$&
${\bf0.8211}$\\
\hline
\hline $(0;*)^-$ & $m$ & $2$& $6$ & $16$
& $52$&
$172$& $586$& $2048$& $7286$\\
\cline{2-10} & $\l$ &{\bf0.7559}& $0.6292$ &
${\bf0.7928}$ & ${\bf0.8048}$& $0.7684$&
${\bf0.8092}$& ${\bf0.8119}$& $-$\\
\hline
\hline
\end{tabular}
\end{center}
\begin{center}
Table 4. The largest eigenvalues in the sectors $(0;0)^+$ and $(0;*)^-$.
\end{center}

We have obtained these numbers by
a direct diagonalization of the transfer matrix in corresponding
subspaces with fixed values $p_2,p_3,\e=\pm1$.
The only missing number is for $N=9$ in the sector $(0;*)^-$.
The size of the transfer matrix becomes too large in this case.

From these results we see that the largest eigenvalue
belongs to the sector $(0;0)^+$  with multiplicity 1 for $N=3l$
and to the sector $(0;*)^-$ with multiplicity 3 when $N$ is not
divisible by 3. Of course, this degeneracy will not affect
the value of $\l$ when the size of the transfer matrix
goes to infinity. Moreover, the largest eigenvalues in the
sectors  $(0;0)^+$ and $(0;*)^-$ are strictly positive.
Also we can see
that when $N$ is divisible by $3$ the value
of $\l$ decreases and when $N$ is not divisible by $3$ the
value of $\l$ increases. One can assume that in the
thermodynamic limit these two sequences
tend to the same number.

Numerical calculations show that for a different (close
to the symmetric point) choice of the spectral parameters
the picture remains essentially the same.
We should note that when $N$ is not multiple of 3 this situation can change
for a different choice of the spectral parameters. However, when $N=3l$
the largest eigenvalue is always in the sector $(0,0)^+$.
Let us restrict our attention only to this case.
As was mentioned above for the  states in the class $\psi^+$
the degree of polynomials $s(p)$ and $\ov s(p)$ is
$k=2N$.

Let us introduce the following parameterization for the parameters
$p_i$, $\ov p_i$ satisfying (\ref{bethe1}-\ref{bethe2})
\be
p_i = -s_i {{e^{-{\it i}\pi/4}}\over{T_1}},\quad
\ov p_i\; =\; {\ov s}_i\; {{e^{{\it i}\pi/4}}\over T_1},\quad
i=1,..,2N,
\label{ptp}
\ee
Numerical analysis shows that $\ov s_i$ and $s_i$ are complex conjugate
for all $i$.

Moreover, for the states in the sector $(0,0)^+$
$2N$ parameters $s_i$  can be divided into
two sets, namely, $\{s_i,s_{i+N}\}$,
$i=1,...,N$ in such a way that
\be
s_i s_{i+N}\;=\;1,\quad i=1,\ldots,N.
\label{cond}
\ee
It is not difficult to see that these equations reduce the system
of equations (\ref{bethe1}-\ref{bethe2}).

 Let us write $s_i$ and $\ov s_i$ as
\be
s_i=e^{x_i+{\it i}y_i},\quad \ov s_i=e^{x_i-{\it i}y_i},\quad i=1,\ldots,N,
\label{xiyi}
\ee
where $x_i$ and $y_i$ are two sets of real numbers.
Let $\zeta$ be a real parameter  such that
\be
T_1=e^{-\zeta}.
\label{T1}
\ee
Then Bethe ansatz equations (\ref{bethe1}-\ref{bethe2})
can be rewritten in the following form
\be
{\Biggl[
{
{\ds\cosh{(x_i + {\it i} y_i \pm {\it i}{\pi\over6})}\;+\;
\cosh{(\zeta\;-\;{\it i}{\pi\over12})}}
\over
{\ds\cosh{(x_i + {\it i} y_i \pm {\it i}{\pi\over6})}\;+\;
\cosh{(\zeta-{\it i}{5\pi\over12})}}
}
\Biggr]}^N=-\prod_{j=1}^N
{
{\ds\cosh{(x_i + {\it i}y_i \pm {\it i}{\pi\over6})}\;-\;
\cosh{(x_j - {\it i} y_j)}}
\over
{\ds\cosh{(x_i + {\it i}y_i \pm {\it i}{\pi\over6})}\;+\;
\cosh{(x_j - {\it i} y_j)}}
}
\label{bethe}
\ee
One has to note that the equations (\ref{bethe}) should be valid for both signs
simultaneously. Of course, its complex conjugations should be valid as well.
So, in fact we have four sets of $N$ equations.
We can easily see that the transformation $(x_i,y_i)\rightarrow(-x_i,-y_i)$
is the symmetry of (\ref{bethe}).

\section{ The two-line hypothesis and the thermodynamic limit}

Now let us consider  the structure of the solutions to the
Bethe ansatz equations (\ref{bethe}) corresponding to the largest eigenvalues
in regime I when $N=3l$.
In Fig.~4
we show the numerical data of the solution to the equations
(\ref{bethe}) for $N=3l$ and $l=1,\ldots,5$

\begin{picture}(1200,500)
\put(90,-800){
\begin{picture}(1100,420)
\thicklines
\put(700,600){\vector(0,1){670}}
\multiput(0,630)(0,120){5}{\vector(1,0){1050}}

\put(30,520){Fig. 4. $\quad$ The structure of Bethe ansatz solutions for
$(x_i,y_i)$ in regime I}
\multiput(1040,1120)(0,-120){5}{$x$}
\put(715,1275){$y$}

\put(1010,600){\small $N=15$}
\put(1010,720){\small $N=12$}
\put(1010,840){\small $N=9$}
\put(1010,960){\small $N=6$}
\put(1010,1080){\small $N=3$}

\put(700,639.163){\line(1,0){8}}
\put(710,650){\small ${{\pi}\over{12}}$}

\put(705,715.793){\small ${{11}\over{12}}\pi$}

\put(700,759.163){\line(1,0){8}}
\put(710,768){\small ${{\pi}\over{12}}$}

\put(700,850.793){\line(1,0){8}}
\put(710,830){\small ${{11}\over{12}}\pi$}

\put(700,879.163){\line(-1,0){8}}
\put(658,888){\small ${{\pi}\over{12}}$}

\put(700,970.793){\line(-1,0){8}}
\put(640,953){\small ${{11}\over{12}}\pi$}

\put(700,999.163){\line(1,0){8}}
\put(710,1008){\small ${{\pi}\over{12}}$}

\put(700,1090.79){\line(-1,0){8}}
\put(640,1072){\small ${{11}\over{12}}\pi$}

\put(700,1119.16){\line(-1,0){8}}
\put(658,1128){\small ${{\pi}\over{12}}$}

\put(700,1210.79){\line(-1,0){8}}
\put(640,1193){\small ${{11}\over{12}}\pi$}

\put(381.174,1208.69){\circle*{5}}
\put(594.757,1209.46){\circle*{5}}
\put(481.742,1122.86){\circle*{5}}

\put(269.922,1088.73){\circle*{5}}
\put(436.254,1090.16){\circle*{5}}
\put(544.755,1090.43){\circle*{5}}
\put(709.929,1090.75){\circle*{5}}
\put(317.857,1001.91){\circle*{5}}
\put(649.487,999.524){\circle*{5}}

\put(204.191,968.688){\circle*{5}}
\put(370.255,970.224){\circle*{5}}
\put(456.158,970.527){\circle*{5}}
\put(527.519,970.653){\circle*{5}}
\put(612.829,970.761){\circle*{5}}
\put(779.333,970.794){\circle*{5}}
\put(235.828,881.706){\circle*{5}}
\put(486.21,879.814){\circle*{5}}
\put(744.748,879.085){\circle*{5}}

\put(156.83,848.66){\circle*{5}}
\put(325.075,850.24){\circle*{5}}
\put(406.896,850.555){\circle*{5}}
\put(465.942,850.675){\circle*{5}}
\put(519.468,850.741){\circle*{5}}
\put(578.297,850.784){\circle*{5}}
\put(659.482,850.799){\circle*{5}}
\put(827.563,850.793){\circle*{5}}
\put(180.473,761.621){\circle*{5}}
\put(407.187,759.797){\circle*{5}}
\put(570.767,759.167){\circle*{5}}
\put(801.655,759.167){\circle*{5}}

\put(119.6927615969970866,728.63956970358747591){\circle*{5}}
\put(290.0216284190285676,730.24439727929926585){\circle*{5}}
\put(371.0468897286728690,730.56690912819110910){\circle*{5}}
\put(426.39805421339884821,730.68602448601870155){\circle*{5}}
\put(471.92235364528443669,730.74404217228814300){\circle*{5}}
\put(514.69959598433057530,730.77644825389993911){\circle*{5}}
\put(560.05773749036489017,730.79384789165907274){\circle*{5}}
\put(614.96426212212945331,730.79793220722128869){\circle*{5}}
\put(694.976680151454003915,730.79341074135456425){\circle*{5}}
\put(865.06517355654065644,730.79276430267224324){\circle*{5}}
\put(138.5754233246897658,641.574398180091637091){\circle*{5}}
\put(354.1024384146762351,639.780481886589767867){\circle*{5}}
\put(490.27932845348184367,639.260506259516900480){\circle*{5}}
\put(627.66333639396182293,639.125541688593916795){\circle*{5}}
\put(844.52493471633580889,639.163077556251610759){\circle*{5}}

\end{picture}}
\end{picture}

\vspace{4cm}

All roots  $(x_i,y_i)$, $i=1,\ldots,N$  split into two sets:
\be
y_i\approx{11\pi\over12},\>\>\> i=1,\ldots,{2N\over3}; \quad
y_i\approx{\pi\over12},\>\>\> i={2N\over3}+1,\ldots,N.
\ee
For finite $N$ these two sets of roots form two curves in
the complex plane. However, for large $N$ these
curves tend to two straight lines: $y={11\pi\over12}$ and $y={\pi\over}$.
Therefore, we expect that  in the thermodynamic
limit $N\rightarrow\infty$ all solutions distribute on two lines
with the imaginary parts $11\pi/12$ and $\pi/12$.

According to a standard method (see, for example,
\cite{BaxB}) let us assume that in the limit $N\rightarrow\infty$
the real parts $x_i$ are distributed
along the whole real axis
with two different densities $\rho_-$ and $\rho_+$
for the first and the second lines respectively.
Let  $N\rho_-(k)dk$ and $N\rho_+(k)dk$ be the
number of the solutions $x_i$ with the imaginary part $y_i=11\pi/12$ and $y_i=\pi/12$
in the interval $k,k+dk$.

To write down the equations on the densities $\rho_+$ and $\rho_-$ we
need the Bethe equations (\ref{bethe}) in the logarithmic form.
Here one should control the phases coming from logarithms.
Taking  logarithm of
(\ref{bethe}) and making  appropriate insertions of signs into
the arguments of logarithms
we can  achieve the  equidistant phase distribution  for both lines.
Then substituting the finite sums by integrals with the corresponding densities
we arrive after some algebra at the following pair of the integral equations
\beq
&{\ds \ln\biggl[-{{\sinh({{k+\zeta}\over 2}-{{{\it i}}\pi\over 6})
\sinh({{k-\zeta}\over 2}-{{{\it i}}\pi\over {12}})}\over
{\sinh({{k+\zeta}\over 2}-{{{\it i}}\pi\over 3})
\sinh({{k-\zeta}\over 2}+{{{\it i}}\pi\over {12}})}} \biggr]+
i\pi\Bigl[{2\over3}-2\int_{-\infty}^k dk'\r_-(k')\Bigr]}
&\nonumber\\
&{\ds -\int^{\infty}_{-\infty}dk'\r_-(k')\ln\biggl\{-
\tanh\Bigl[{k-k'\over2}-{\it i}{\pi\over6}\Bigr]
\tanh\Bigl[{k+k'\over2}-{\it i}{\pi\over12}\Bigr]
\biggl\} }&\nonumber\\
&{\ds +\int^{\infty}_{-\infty}dk'\r_+(k')\ln\biggl\{-
\tanh\Bigl[{k-k'\over2}-{\it i}{\pi\over12}\Bigr]
\tanh\Bigl[{k+k'\over2}-{\it i}{\pi\over6}\Bigr]
\biggl\}=0\phantom{-..} }&
\label{intbethe1}\eeq
\beq
&{\ds \ln\biggl[{{\cosh({{k+\zeta}\over 2}+{{{\it i}}\pi\over {12}})
\cosh({{k-\zeta}\over 2}+{{{\it i}}\pi\over {6}})}\over
{\cosh({{k+\zeta}\over 2}-{{{\it i}}\pi\over {12}})
\cosh({{k-\zeta}\over 2}+{{{\it i}}\pi\over {3}})}} \biggr]-
i\pi\Bigl[{1\over3}-2\int_{-\infty}^kdk'\r_+(k')\Bigr] }
&\nonumber\\
&{\ds -\int^{\infty}_{-\infty}dk'\r_+(k')\ln\biggl\{-
\tanh\Bigl[{k-k'\over2}+{\it i}{\pi\over6}\Bigr]
\tanh\Bigl[{k+k'\over2}+{\it i}{\pi\over12}\Bigr]
\biggl\} }&\nonumber\\
&{\ds +\int^{\infty}_{-\infty}dk'\r_-(k')\ln\biggl\{-
\tanh\Bigl[{k-k'\over2}+{\it i}{\pi\over12}\Bigr]
\tan\Bigl[{k+k'\over2}+{\it i}{\pi\over6}\Bigr]
\biggl\}=0\phantom{-..} }&
\label{intbethe2}
\eeq
with   normalization conditions
\be
\int_{-\infty}^\infty dk\r_+(k)={1\over3},\quad
\int_{-\infty}^\infty dk\r_-(k)={2\over3}.
\label{norm}
\ee
Then differentiating both equations
by $k$ and making the Fourier transform we obtain the solution to these
equations
\be
\rho_{\pm}(k)={{\sqrt{3}/\pi}\over{2\cosh{[2(\zeta-k)]}\pm 1}}.
\label{resrho}
\ee
Substituting these formulae into the initial integral equations
(\ref{intbethe1}-\ref{intbethe2}) one can check that they are really satisfied.

In Fig.~5, 6 we plot the value ${1\over N}{1\over{x_{i+1}-x_{i}}}$
as a function of $x_i$
with $y_i\approx {11\pi\over12}$, $y_i\approx {\pi\over12}$ for $N=90$ and
the functions $\rho_-(x)$, $\rho_+(x)$  at $\zeta\approx-0.58666$.
From these figures one can see that discrete distributions are very close to
its continuous limits.

\newpage

\begin{picture}(600,300)
\put(-50,-250)
{
\par\noindent
 \centerline{\epsfxsize=4.5in\epsfbox{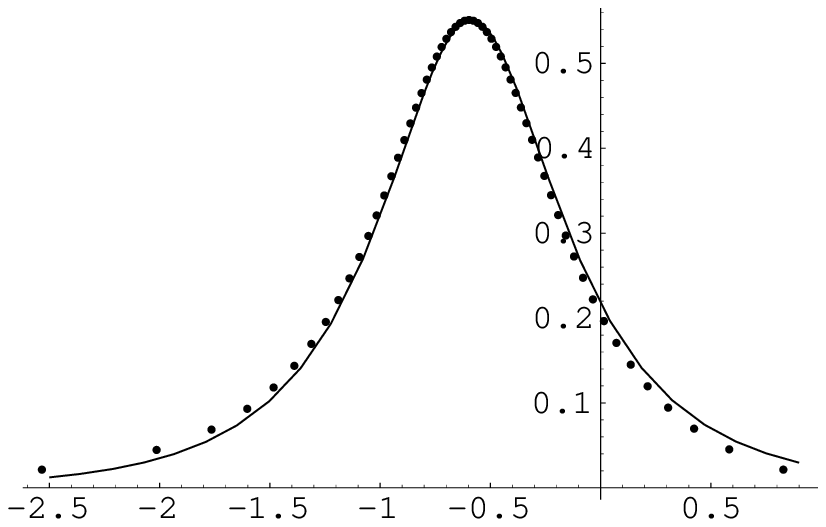}}
\par\noindent
}
\put(220,285){\line(1,0){80}}
\put(340,280){$\rho_-(x)$}
\put(250,156){\circle*{12}}
\put(300,150){$\ds\small{1\over N}{1\over{x_{i+1}-x_{i}}}$}
\put(250,50){$N=90,\quad 1\le i\le59$}
\put(1180,-230){$\large x$}
\put(-50,-300)
{
\noindent
\centerline{{\bf Fig. 5.}
 {\bf ~Discrete distribution of $x_i$ with $y_i={{11\pi}\over{12}}$ and the
 function
 $\rho_-(x)$.}}
}
\end{picture}
\vspace{5cm}

\begin{picture}(600,300)
\put(-50,-250)
{
\par\noindent
 \centerline{\epsfxsize=4.5in\epsfbox{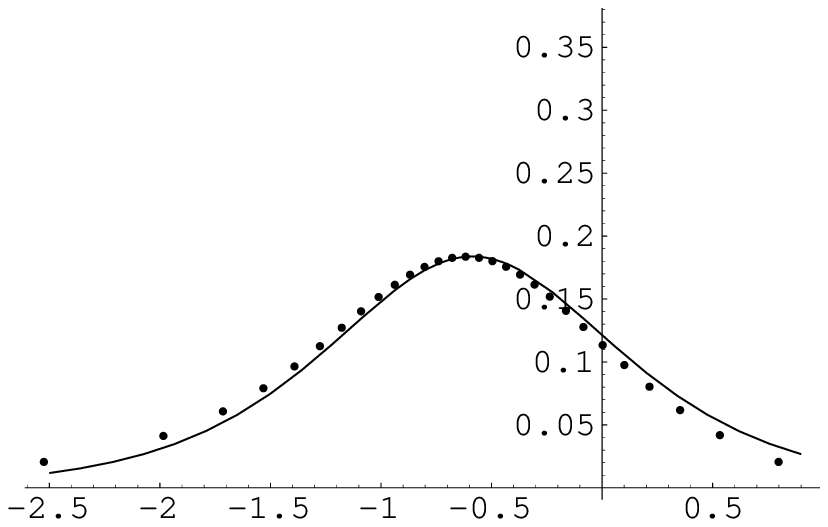}}
\par\noindent
}
\put(220,285){\line(1,0){80}}
\put(340,280){$\rho_+(x)$}
\put(250,156){\circle*{12}}
\put(300,150){$\ds\small{1\over N}{1\over{x_{i+1}-x_{i}}}$}
\put(250,50){$N=90,\quad 60\le i\le89$}
\put(1180,-230){$\large x$}
\put(-50,-300)
{
\noindent
\centerline{{\bf Fig. 6.}
 {\bf ~Discrete distribution of $x_i$ with $y_i={{\pi}\over{12}}$ and   the
 function
 $\rho_+(x)$.}}
}
\end{picture}
\vspace{4cm}

\section{Calculation of the partition function per site $\kappa$}

In this section we calculate the partition function
$\kappa$ given by (\ref{kappa2}) in the thermodynamic limit for the case of regime I.
First of all, using the formulae (\ref{tt},\ref{sts},\ref{kappa2},\ref{a2N}) we obtain
for the finite $N$
\be
\ln{(\kappa/2)}=-{{2\ln{2}}\over3}+K_1+K_2+K_3+K_4+{{\ln{4}}\over{3N}}-
{1\over 3}\ln{[(p+q)(p+q')(p'-q)(p'+q')]}+{1\over 3}S_1+{1\over 3}S_2
\label{kappa3}
\ee
where
\be
S_1={1\over{N}}\sum_{i=1}^{N}\ln{(p-p_i)}+{1\over{N}}\sum_{i=1}^{N}\ln{(p-p_{i+N})},\quad
S_2={1\over{N}}\sum_{i=1}^{N}\ln{(p'-\ov p_i)}+
{1\over{N}}\sum_{i=1}^{N}\ln{(p'-\ov p_{i+N})}.
\label{S12}
\ee

Now let us consider the thermodynamic limit and
replace all sums in (\ref{S12}) by integrals with corresponding
densities.
As above we can face  the problem of  sign choice in arguments of
logarithms in integrals.
One can choose these signs comparing each integral with a
numerical value of the corresponding sum for finite $N$.
Proceeding this way we obtain
$$
s_1=\lim_{N\to\infty}S_1=
\int_{-\infty}^{\infty}dk\rho_-(k)\ln{(p+e^{\zeta-{\it i}\pi/4}
e^{k+{\it i}11\pi/12})}+
\int_{-\infty}^{\infty}dk\rho_+(k)\ln{(p+e^{\zeta-{\it i}\pi/4}
e^{k+{\it i}\pi/12})}+
$$
\be
\int_{-\infty}^{\infty}dk\rho_-(k)\ln{(p+e^{\zeta-{\it i}\pi/4}
e^{-k-{\it i}11\pi/12})}+
\int_{-\infty}^{\infty}dk\rho_+(k)\ln{(p+e^{\zeta-{\it i}\pi/4}
e^{-k-{\it i}\pi/12})},
\label{S1}
\ee
\vspace{0.3cm}
$$
s_2=\lim_{N\to\infty}S_2=
\int_{-\infty}^{\infty}dk\rho_-(k)\ln{(p'-e^{\zeta+{\it i}\pi/4}
e^{k-{\it i}11\pi/12})}+
\int_{-\infty}^{\infty}dk\rho_+(k)\ln{(-p'+e^{\zeta+{\it i}\pi/4}
e^{k-{\it i}\pi/12})}+
$$
\be
\int_{-\infty}^{\infty}dk\rho_-(k)\ln{(p'-e^{\zeta+{\it i}\pi/4}
e^{-k+{\it i}11\pi/12})}+
\int_{-\infty}^{\infty}dk\rho_+(k)\ln{(-p'+e^{\zeta+{\it i}\pi/4}
e^{-k+{\it i}\pi/12})}.
\label{S2}
\ee
Now using the solution for the densities $\rho_+$ and $\rho_-$ (\ref{resrho}),
making some shifts in the integrals and taking into account
(\ref{qq'}) we come to the following formulae for $s_1$ and $s_2$
$$
s_1=
{{\sqrt{3}}\over{\pi}}\int_{-\infty}^{\infty}dk
{{\ln{[v_4\,(1-{v_3}^{-1}e^{k+i\pi/6})]}}\over{2\cosh{2k}-1}}
+
{{\sqrt{3}}\over{\pi}}\int_{-\infty}^{\infty}dk
{{\ln{[v_4\,(1+{v_3}^{-1}e^{k+i\pi/3})]}}\over{2\cosh{2k}+1}}
+
$$
\be
{{\sqrt{3}}\over{\pi}}\int_{-\infty}^{\infty}dk
{{\ln{[-e^{-i\pi/6}\,(1- v_4 e^{k+i\pi/6})]}}\over{2\cosh{2k}-1}}
+
{{\sqrt{3}}\over{\pi}}\int_{-\infty}^{\infty}dk
{{\ln{[e^{-i\pi/3}\,(1 + v_4 e^{k+i\pi/3})]}}\over{2\cosh{2k}+1}} ,
\label{S1a}
\ee

\vspace{0.4cm}

$$
s_2=
{{\sqrt{3}}\over{\pi}}\int_{-\infty}^{\infty}dk
{{\ln{[{\ds
{{v_2}\over{v_1}}\,e^{-i\pi/6}\,(1-{v_2}^{-1}e^{k+i\pi/6})}]}}\over{2\cosh{2k}-1}}
+
{{\sqrt{3}}\over{\pi}}\int_{-\infty}^{\infty}dk
{{\ln{[{\ds
{{v_2}\over{v_1}}\,e^{-i\pi/3}\,(1+{v_2}^{-1}e^{k+i\pi/3})}]}}\over{2\cosh{2k}+1}}
+
$$
\be
{{\sqrt{3}}\over{\pi}}\int_{-\infty}^{\infty}dk
{{\ln{[{\ds
-{1\over{v_1}}\,(1- v_1 e^{k+i\pi/6})}]}}\over{2\cosh{2k}-1}}
+
{{\sqrt{3}}\over{\pi}}\int_{-\infty}^{\infty}dk
{{\ln{[{\ds
{1\over{v_1}}\,(1 + v_1 e^{k+i\pi/3})}]}}\over{2\cosh{2k}+1}}.
\label{S2a}
\ee

\vspace{0.4cm}
Using  formulas (\ref{K}) and (\ref{qq'})
we rewrite (\ref{kappa3}) after some algebra  as
\be
\ln{(\kappa/2)}=F(v_1)+F(v_2^{-1})+F(v_3^{-1})+F(v_4)
\label{kappa4}
\ee
for
\be
0<\th_i<\pi,\quad 0<a_i<\pi
\label{condpar}
\ee
where
\be
F(v)=-{{{\it i}\pi}\over{27}}-
{1\over{12}}\ln{(1+v)}-{1\over{4}}\ln{(1-v)}+{1\over 3}I(v)
\label{FF}
\ee
and
\be
I(v)=I_+(v\,e^{{\it i}\pi/3})+I_-(-v\, e^{{\it i}\pi/6})
\label{I}
\ee
with
\be
I_{\pm}(z)={{\sqrt{3}}\over{\pi}}\int_{-\infty}^{\infty}{{dk}\over{2\cosh{2k}\pm 1}}
\ln{(1+z\,e^k)}
\label{Ipm}
\ee

Taking into account that $v_2={v_4^*}, v_3={v_1^*}$ and
\be
I(v^{-1})={\tilde I}(v)+\ln(v^{-1})+{\it i}\phi(v)
\label{ItildeI}
\ee
where
\be
{\tilde I}(v)=I_+(v\,e^{-{\it i}\pi/3})+I_-(-v\, e^{-{\it i}\pi/6})
\label{tildeI}
\ee
and
\be
\phi(v)=\cases{
{\ds  -{{4\pi}\over{9}} }& if ${\ds -{{2\pi}\over 3}<\arg{v}<{{\pi}\over 6}}$ \cr
{\ds {{8\pi}\over{9}}} & if ${\ds {{\pi}\over 6}<\arg{v}<\pi}$ \cr
{\ds  -{{\pi}\over{9}} }& if ${\ds \pi<\arg{v}<{{4\pi}\over 3}}$ \cr
}
\label{phi}
\ee
one can check that the pure imaginary term $-{{{\it i}\pi}\over{27}}$ in (\ref{FF})
compensates the imaginary part in the whole expression
(\ref{kappa4}) for all values of parameters in the domain (\ref{condpar}).
So the final expression (\ref{kappa4}) is real as it should be. We can also
note that the form of (\ref{kappa4}) is similar to the Baxter's formula (\ref{kappa})
for $n=3$ with the function $J_3$ instead of $J$ defined by (\ref{J_n}) and
(\ref{J_nhat}).
In terms of the variables $v_i$ the formula (\ref{kappa}) would be
$\Phi(v_1)+\Phi(-v_2^{-1})+\Phi(v_3^{-1})+\Phi(-v_4)$ with
$\Phi(-{\it i}e^{-{\it i}x})=J_3(x)$. A difference with (\ref{kappa4}) is the
presence of the additional signs before $v_2^{-1}$ and $v_4$. Moreover, the function
$F$ given by (\ref{FF}) is  different in comparison with
the function $\Phi$.

Now using some integral formulae  from \cite{GR} and
taking into account the cuts of the integrals (\ref{Ipm}) we can obtain
the following expression for the combination (\ref{I})
\be
\label{Ia}
I(z)=
\cases{
I^{(0)}(z) & ${\ds -{{\pi}\over 6}<\arg{z}<{{\pi}\over 6}}$\cr
I^{(0)}(z)-{1\over 2}\ln{(1-\om z^2)}&
${\ds  {{\pi}\over 6}<\arg{z}<{{\pi}\over 2}}$\cr
I^{(0)}(z)-{1\over 2}\ln{(1-\om z^2)}+\ln{(1-z^2)}
& ${\ds \> \> \> \> {{\pi}\over 2}<\arg{z}<{{2\pi}\over 3}}$\cr
I^{(0)}(z)-{1\over 2}\ln{(1-\om z^2)}+\ln{(1-\om^{-1}z^2)}
& ${\ds {{2\pi}\over 3}<\arg{z}<{{5\pi}\over 6}}$\cr
I^{(0)}(z)-{1\over 2}\ln{(1-\om z^2)}+{1\over 2}\ln{(1-\om^{-1}z^2)}
& ${\ds  \> \> \> \> {{5\pi}\over 6}<\arg{z}<{{7\pi}\over 6}}$\cr
I^{(0)}(z)-\ln{(1-\om z^2)}+{1\over 2}\ln{(1-\om^{-1}z^2)}\;
&  ${\ds {{7\pi}\over 6}<\arg{z}<{{3\pi}\over 2}}$\cr
I^{(0)}(z)-\ln{(1-\om z^2)}+{1\over 2}\ln{(1-\om^{-1}z^2)}+\ln{(1-z^2)}
&  ${\ds  \> \> \> \> {{3\pi}\over 2}<\arg{z}<{{11\pi}\over 6}}$\cr }
\ee
Here $I^{(0)}(z)$ corresponds to the principal branch of the integrals in (\ref{I})
$$
I^{(0)}(z)=\ln{(1+\om^{-1}z)}+{1\over 2}\ln{(1-\om^{-1}z)}-{1\over 2}\ln{(1-\om z)}+
{{{\it i}}\over{4\pi}}[2{\mbox{Li}}_2(1-z^2)-{\mbox{Li}}_2(1-\om z^2)
-{\mbox{Li}}_2(1-\om^{-1} z^2)]
$$
\be
\mbox{for} -\pi/6<\arg{z}<\pi/6
\label{I0}
\ee
where $\om=e^{{\it i}2\pi/3}$ and ${\mbox{Li}}_2(z)$ is Euler dilogarithm
\be
{\mbox{Li}}_2(z)=\sum_{n=1}^{\infty}{{z^n}\over{n^2}}=-\int_0^z dt\;{{\ln{(1-t)}}\over t}.
\label{Euler}
\ee

Now we can  compare
the asymptotics of $\ln(\kappa/2)$ with our result. Namely, for the
parameters $\th_1=2.541,\th_2=1.1,\th_3=1.3$ we get from the formula
(\ref{kappa4}) approximately $\ln(\kappa/2)=0.2224078...$.
Fig.~7 shows a dependence of $\ln(\kappa/2)$ on $N$ for this choice
of the spectral parameters.


\begin{picture}(1150,380)
\put(100,20){
\begin{picture}(1100,350)
\thicklines
\put(80,0){\vector(0,1){300}}
\put(80,0){\vector(1,0){1020}}

\put(120,300){$\ln(\kappa/2)$}

\put(900,250){$N=3k$}

\put(1050,-40){ $N$}

\put(150,-100){Fig. 7. $\quad$ Logarithm of the partition function in regime I}

\put(-20,-10){$0.218$}

\put(-20,201){$0.242$}
\put(80,206.152){\line(-1,0){5}}
\put(-20,113){$0.232$}
\put(80,118.225){\line(-1,0){5}}
\put(-20,35){$0.2226$}
\put(80,40.1945){\line(-1,0){5}}

\put(90,-35){ $3$}
\put(120,-35){ $6$}
\put(150,-35){ $9$}
\put(350,-35){ $30$}
\put(650,-35){ $60$}
\put(950,-35){ $90$}

\put(140,206.152){\circle*{5}}
\put(170,118.225){\circle*{5}}
\put(200,84.518){\circle*{5}}
\put(230,68.247){\circle*{5}}
\put(260,59.220){\circle*{5}}
\put(290,53.716){\circle*{5}}
\put(320,50.121){\circle*{5}}
\put(350,47.647){\circle*{5}}
\put(380,45.87498){\circle*{5}}
\put(410,44.5617){\circle*{5}}
\put(440,43.562){\circle*{5}}
\put(470,42.784){\circle*{5}}
\put(500,42.166){\circle*{5}}
\put(530,41.668){\circle*{5}}
\put(560,41.260){\circle*{5}}
\put(590,40.922){\circle*{5}}
\put(620,40.6389){\circle*{5}}
\put(650,40.399){\circle*{5}}
\put(680,40.1945){\circle*{5}}
\put(710,40.0184){\circle*{5}}
\put(740,39.86577){\circle*{5}}
\put(770,39.7326){\circle*{5}}
\put(800,39.6157){\circle*{5}}
\put(830,39.5126){\circle*{5}}
\put(860,39.421){\circle*{5}}
\put(890,39.339689){\circle*{5}}
\put(920,39.266789){\circle*{5}}
\put(950,39.20130){\circle*{5}}
\put(980,39.142255){\circle*{5}}

\put(110,-4){\line(0,1){4}}
\put(140,-4){\line(0,1){4}}
\put(170,-4){\line(0,1){4}}
\put(200,-4){\line(0,1){4}}
\put(230,-4){\line(0,1){4}}
\put(260,-4){\line(0,1){4}}
\put(290,-4){\line(0,1){4}}
\put(320,-4){\line(0,1){4}}
\put(350,-4){\line(0,1){4}}
\put(380,-10){\line(0,1){10}}
\put(410,-4){\line(0,1){4}}
\put(440,-4){\line(0,1){4}}
\put(470,-4){\line(0,1){4}}
\put(500,-4){\line(0,1){4}}
\put(530,-4){\line(0,1){4}}
\put(560,-4){\line(0,1){4}}
\put(590,-4){\line(0,1){4}}
\put(620,-4){\line(0,1){4}}
\put(650,-4){\line(0,1){4}}
\put(680,-10){\line(0,1){10}}
\put(710,-4){\line(0,1){4}}
\put(740,-4){\line(0,1){4}}
\put(770,-4){\line(0,1){4}}
\put(800,-4){\line(0,1){4}}
\put(830,-4){\line(0,1){4}}
\put(860,-4){\line(0,1){4}}
\put(890,-4){\line(0,1){4}}
\put(920,-4){\line(0,1){4}}
\put(950,-4){\line(0,1){4}}
\put(980,-10){\line(0,1){10}}

\end{picture}}
\end{picture}

\vspace{2cm}

From Fig.~7 we can also see
that for $N>30$ the two digits in $\ln(\kappa/2)\approx0.22$  are fixed. It is
in an agreement with its limiting value $0.2224078...$

In the end of this section
let us consider an isotropic case of the Zamolodchikov model when
all spherical angles and sides $\theta_i=a_i=\pi/2$.
Then $v_1=1/v_3=-e^{{\it i}\pi/4}$ and $v_4=1/v_2=e^{{\it i}\pi/4}$.
Using the formula (\ref{Ia}) one can get
\be
I(e^{{\it i}\pi/4})+I(-e^{{\it i}\pi/4})=
{{5 G}\over{3\pi}}+{1\over 3} \ln{(\sqrt{3}+1)} + {{\ln{2}}\over{12}}-
{{{\it i}\pi}\over{36}}
\label{Isym}
\ee
where as in  (\ref{kappaisot}), $G$ is Catalan constant.
Substituting it into the formula (\ref{kappa4}) we obtain
the answer for the logarithm of the partition function in the isotropic case:
\be
\ln{(\kappa/2)}=
{{10 G}\over{9\pi}}+{2\over 9} \ln{(\sqrt{3}+1)} - {{5\ln{2}}\over{18}}
\approx 0.354760685.
\label{kappasym}
\ee

\section{The functional relations in the thermodynamic limit}

In the previous sections we have obtained the integral equations
(\ref{intbethe1}-\ref{intbethe2}) using the two-line hypothesis
resulting from our numerical analysis of the solutions to the Bethe
ansatz equations. Here we would like to discuss a compatibility
of the results obtained above with the functional relations (4.1-4.4)
of \cite{BM2}
for the transfer matrix eigenvalues $t(p;q,q')$ and $\ov t(p;q,q')$ or
(4.9-4.11) of \cite{BM2} for  $s(p;q,q')$ and $\ov s(p;q,q')$.
Let us remind the reader that the functional relations for
$s(p;q,q')$ and $\ov s(p;q,q')$ look as follows
$$
{\ov s(p)}\;\;s(p)\;\;{\ov s(-p)}\;\;s(-\om p)\,=\,
$$
$$
{\l}_0^N\;{\ov s(p)}\,s(-\om p)\,+\,
{\l}_1^N\;{\ov s(-p)}\,s(-\om p)\,+\,
{\l}_2^N\;{\ov s(p)}\,s(\om p)\,+\,
{\l}_3^N\;{\ov s(-p)}\,s(\om p)
$$
\be
\quad
\label{fes}
\ee
and
$$
{\ov s(-\om p)}\;\;s(-p)\;\;{\ov s(p)}\;\;s(p)\,=\,
$$
$$
{\l'}_0^N\;{\ov s(-\om p)}\,s(p)\,+\,
{\l'}_1^N\;{\ov s(-\om p)}\,s(-p)\,+\,
{\l'}_2^N\;{\ov s(\om p)}\,s(p)\,+\,
{\l'}_3^N\;{\ov s(\om p)}\,s(-p)
$$
\be
\quad
\label{fes1}
\ee
where for simplicity a dependence on $q,q'$ has been omitted and
$$
{\l}_0\;=\;(p\,+\om\,q)\;(p\,+\om^{-1}\,q)\;(p\,+\,q')\;(p\,-\,q')
$$
$$
{\l}_1\;=\;(p\,+\om\,q')\;(p\,+\om^{-1}\,q')\;(p\,+\,q)\;(p\,-\,q)
$$
$$
{\l}_2\;=\;(p\,-\,q)\;(p\,+\om^{-1}\,q)\;(p\,-\,\om q')\;(p\,-\,q')
$$
\be
{\l}_3\;=\;(p\,-\,q')\;(p\,+\om^{-1}\,q')\;(p\,-\,\om q)\;(p\,-\,q)
\label{l}
\ee
with $\l'_i$ being obtained from $\l_i$ by the substitution
$q\rightarrow -q$. The relations (\ref{fes}) and (\ref{fes1}) played
a key role for a derivation of the Bethe ansatz equations  (\ref{bethe1}) and
(\ref{bethe2}).

Denote
\be
u(p;q,q')=\lim_{N\to\infty}{1\over N}\ln{{s(p;q,q')\over a_{2N}(q,q')}} ,\quad
\ov u(p;q,q')=\lim_{N\to\infty}{1\over N}\ln{{\ov s(p;q,q')\over \ov a_{2N}(q,q')}}.
\label{uup}
\ee
Using (\ref{a2N}) and the fact that the relation (\ref{fes}) depends only on the product
of $s(p)$ and $\ov s(p)$ we note that $ a_{2N}(q,q')$ and $\ov a_{2N}(q,q')$
disappear from (\ref{fes}).

Now let us start with the expressions (\ref{S1}) and (\ref{S2})
for $s_1$ and $s_2$. In fact, $s_1=u(p)$ and $s_2=u(p')$ in  (\ref{uup}),
but now we consider them as functions of arbitrary arguments (we fixed
and omitted a dependence on $q, q'$).

It is convenient to extract $\ln{p}$ terms from both
expressions (\ref{S1}), (\ref{S2}).
Due to the fact that $\int_{-\infty}^{\infty}dk\rho_+(k)+
\int_{-\infty}^{\infty}dk\rho_-(k)=1$, it will give $2\ln{p}$  for the
$u(p)$ and for $\ov u(p)$. Then taking into account the definitions
(\ref{Ipm}) up to $2\ln{p}$ we get
\be
u(p)\sim I_+({z\over p}e^{-{\it i}\pi/6})+I_+({1\over p}e^{-{\it i}\pi/3})+
I_-(-{z\over p}e^{-{\it i}\pi/3})+I_-(-{1\over p}e^{-{\it i}\pi/6})
\label{lns}
\ee
\be
\ov u(p)\;\sim \;
I_+(-{z\over p}e^{{\it i}\pi/6})\;+\;I_+(-{1\over p}e^{{\it i}\pi/3})\;+\;
I_-({z\over p}e^{{\it i}\pi/3})\;+\;I_-({1\over p}e^{{\it i}\pi/6})
\label{lnts}
\ee
where for simplicity we  denoted $z=1/{T_1^2}=e^{2\zeta}$ being not
confused with another $z$ defined in (\ref{zT}).

Now we can consider logarithms divided by $N$
of the lhs of the functional relations (\ref{fes}-\ref{fes1}) using the
formulae (\ref{lns}) and (\ref{lnts}). The same thing can be done for each of the four
terms in the rhs of the relations (\ref{fes}-\ref{fes1}).
After this we can study a dominance of different terms in the rhs of (\ref{fes}-\ref{fes1})
in the complex plane of the variable $p$.
In Fig.~8 we show a behaviour of the logarithms  of the absolute values  for each of the four
terms in the rhs of (\ref{fes}) as a functions of $\arg{p}$ at $N=60$
dividing the result by $N$.

\newpage

\begin{picture}(600,300)
\put(-50,-250)
{
\par\noindent
 \centerline{\epsfxsize=4.5in\epsfbox{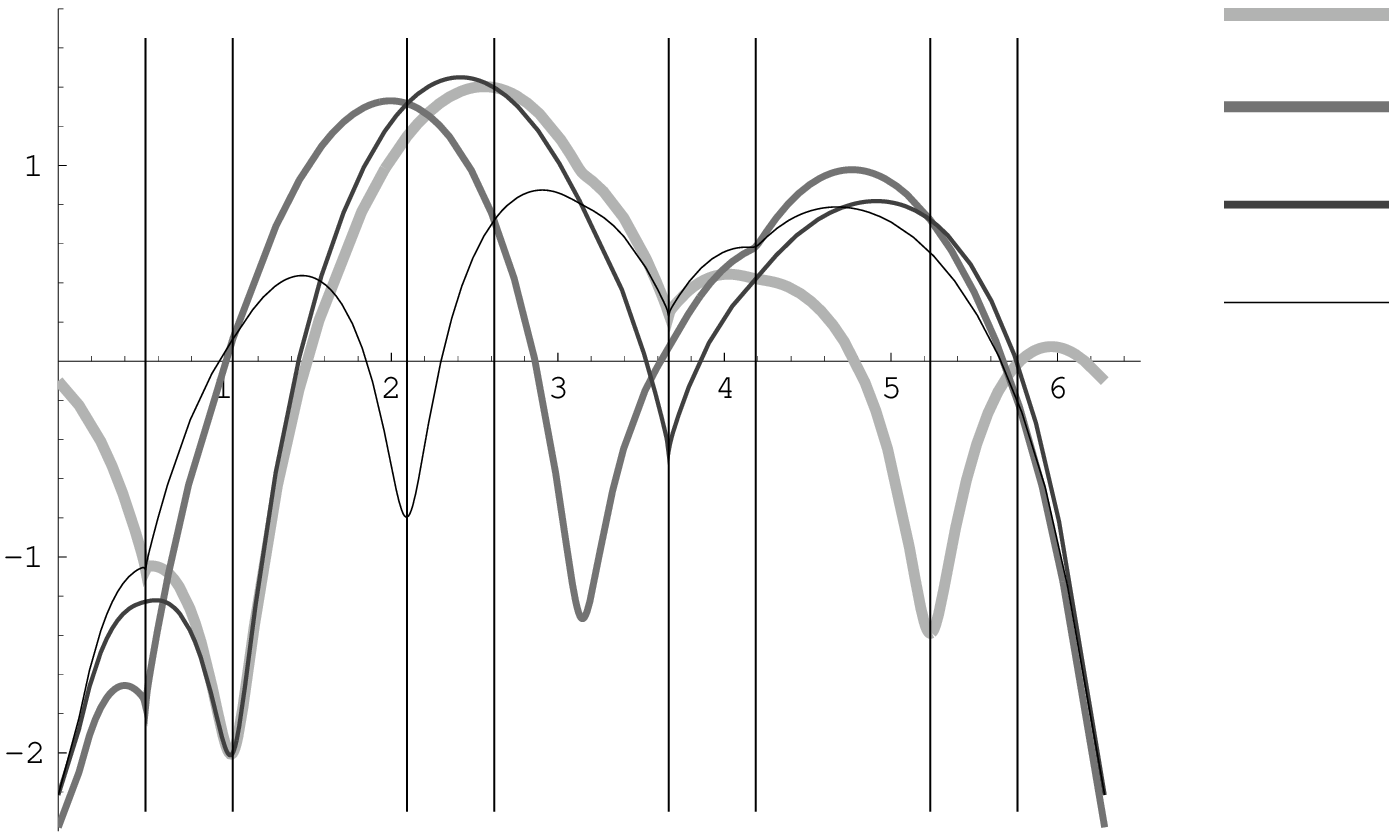}}
\par\noindent
}
\put(315,310){${{\pi}\over 6}$}
\put(373,310){${{\pi}\over 3}$}
\put(485,310){${{2\pi}\over 3}$}
\put(540,310){${{5\pi}\over 6}$}
\put(658,310){${{7\pi}\over 6}$}
\put(715,310){${{4\pi}\over 3}$}
\put(835,310){${{5\pi}\over 3}$}
\put(888,310){${{11\pi}\over 6}$}
\put(990,30){$2\pi$}

\put(1200,285){I term}
\put(1200,225){II term}
\put(1200,160){III term}
\put(1200,92){IV term}


\put(-50,-300)
{
\noindent
\centerline{{\bf Fig. 8.}
 {\bf A behaviour of logarithms of four terms in the rhs of the functional}}
}
\put(-50,-340)
{
\centerline{\bf{
relation (\ref{fes}) within a region $0\le \arg p\le 2\pi$
in regime I}}
}
\end{picture}

\vspace{4.8cm}

From this picture we can divide the complex plane
into the four domains where the different terms dominate. Namely,\\
$\quad$\\
I $\quad$ the first term dominates in the domain I:
\be
-\pi/6<\arg{p}<\pi/6\quad\mbox{and}\quad 5\pi/6<\arg{p}<7\pi/6
\label{domain1}
\ee
II $\quad$  the second term dominates in the domain II:
\be
\pi/3<\arg{p}<2\pi/3\quad\mbox{and}\quad 4\pi/3<\arg{p}<5\pi/3
\label{domain2}
\ee
III $\quad$ the third term dominates in the domain III:
\be
2\pi/3<\arg{p}<5\pi/6\quad\mbox{and}\quad 5\pi/3<\arg{p}<11\pi/6
\label{domain3}
\ee
IV $\quad$ the fourth term dominates in the domain IV:
\be
\pi/6<\arg{p}<\pi/3\quad\mbox{and}\quad 7\pi/6<\arg{p}<4\pi/3.
\label{domain4}
\ee
In Fig.~9 these four domains are shown:

\begin{picture}(800,550)
\put(350,50){
\begin{picture}(700,450)
\thicklines
\put(0,250){\line(1,0){100}}
\put(150,250){\line(1,0){300}}
\put(500,250){\vector(1,0){100}}
\put(300,50){\line(0,1){50}}
\put(300,150){\line(0,1){200}}
\put(300,400){\vector(0,1){50}}
\multiput(300,250)(1,-2){3}{\line(2,1){300}}
\multiput(300,250)(1,2){3}{\line(2,-1){300}}
\multiput(300,250)(2,-1){3}{\line(1,2){100}}
\multiput(300,250)(2,1){3}{\line(-1,2){100}}
\multiput(300,250)(2,1){3}{\line(1,-2){100}}
\multiput(300,250)(2,-1){3}{\line(-1,-2){100}}
\multiput(300,250)(1,2){3}{\line(-2,1){300}}
\multiput(300,250)(1,-2){3}{\line(-2,-1){300}}
\put(470,240){I}
\put(120,240){I}
\put(395,340){IV}
\put(195,140){IV}
\put(290,360){II}
\put(290,110){II}
\put(390,140){III}
\put(190,340){III}

\put(250,-20){\bf Fig. 9.}

\end{picture}}
\end{picture}

So, in the thermodynamic limit for these four domains (\ref{domain1}-\ref{domain4})
we should have four different relations.
Namely, for the region (\ref{domain1}) the first term in the rhs of (\ref{fes})
dominates and we should have the following relation
\be
u(p)+{\ov u(-p)}=\ln{[(p+\om q)(p+\om^{-1} q)(p+q')(p-q')]}.
\label{rel1}
\ee
Extracting the terms like $\ln{p}$ as was described above and using (\ref{qq'}) we get
for the lhs of (\ref{rel1})
$$
I_+({z\over p}e^{-{\it i}\pi/6})+I_+({1\over p}e^{-{\it i}\pi/3})+
I_-(-{z\over p}e^{-{\it i}\pi/3})+I_-(-{1\over p}e^{-{\it i}\pi/6})
$$
\be
I_+({z\over p}e^{{\it i}\pi/6})+I_+({1\over p}e^{{\it i}\pi/3})+
I_-(-{z\over p}e^{{\it i}\pi/3})+I_-(-{1\over p}e^{{\it i}\pi/6}).
\label{lhs1}
\ee
Now using the following relations
\be
I_+(e^{{\it i}\pi/2}w)+I_-(w)=\ln{(1+\, e^{{\it i}\pi/6}w)}
\quad\mbox{for}\quad-\pi<\arg{w}<\pi/2
\label{sum1}
\ee
$$
I_+(e^{{\it i}\pi/2}w)+I_-(w)=\ln{(1- e^{-{\it i}\pi/6}w)}+
\ln{(1+\, e^{-{\it i}\pi/6}w)}-\ln{(1- e^{{\it i}\pi/6}w)}
$$
\be
\quad\mbox{for}\quad\pi/2<\arg{w}<\pi
\label{sum2}
\ee
\be
I_+(e^{-{\it i}\pi/2}w)+I_-(w)=\ln{(1+ e^{-{\it i}\pi/6}w)}
\quad\mbox{for}\quad-\pi/2<\arg{w}<\pi
\label{sum1a}
\ee
$$
I_+(e^{-{\it i}\pi/2}w)+I_-(w)=\ln{(1+ e^{{\it i}\pi/6}w)}+
\ln{(1- e^{{\it i}\pi/6}w)}-\ln{(1- e^{-{\it i}\pi/6}w)}
$$
\be
\quad\mbox{for}\quad\pi<\arg{w}<3\pi/2
\label{sum2a}
\ee
one can get the following result
for the expression (\ref{lhs1}) combining the pairs of
$I_+$ and $I_-$ in accordance with (\ref{sum1}-\ref{sum2a})
$$
\ln{(1+{z\over p}e^{{\it i}\pi/2})}+\ln{(1+{1\over p}e^{{\it i}2\pi/3})}+
\ln{(1-{z\over p}e^{{\it i}\pi/2})}+\ln{(1+{1\over p}e^{-{\it i}2\pi/3})}=
$$
\be
\ln{(1+{{q'}\over p})}+\ln{(1-{{q'}\over p})}+\ln{(1+{{\om}\over p})}+
\ln{(1+{{\om^{-1}}\over p})}.
\label{lhs1a}
\ee
It is easy to see that
the last expression is just the rhs of (\ref{rel1}) after extracting
logarithms of $p$.
For the second domain II (\ref{domain2}) the second term in the rhs of (\ref{fes})
dominates and we should have the following relation
\be
u(p)+{\ov u(p)}=\ln{[(p+\om q')(p+\om^{-1} q')(p+q)(p-q)]}
\label{rel2}
\ee
which can be checked in a similar way as (\ref{rel1}) using relations
(\ref{sum1}-\ref{sum2a}).
For the third domain III (\ref{domain3}) the third term in the rhs of (\ref{fes})
dominates and we have to check a little bit more complicated relation
\be
u(p)+{\ov u(-p)}+u(-\om p)-u(\om p)=
\ln{[(p-q)(p+\om^{-1}q)(p-\om q')(p-q')]}.
\label{rel3}
\ee
In the lhs of (\ref{rel3}) we get
$$
I_+({z\over p}e^{-{\it i}\pi/6})+I_+({1\over p}e^{-{\it i}\pi/3})+
I_-(-{z\over p}e^{-{\it i}\pi/3})+I_-(-{1\over p}e^{-{\it i}\pi/6})+
$$
$$
I_+({z\over p}e^{{\it i}\pi/6})+I_+({1\over p}e^{{\it i}\pi/3})+
I_-(-{z\over p}e^{{\it i}\pi/3})+I_-(-{1\over p}e^{{\it i}\pi/6})+
$$
$$
I_+({z\over p}e^{{\it i}\pi/6})+I_+({1\over p})+
I_-(-{z\over p})+I_-(-{1\over p}e^{{\it i}\pi/6})-
$$
\be
I_+(-{z\over p}e^{{\it i}\pi/6})-I_+(-{1\over p})- I_-({z\over
p})-I_-({1\over p}e^{{\it i}\pi/6}). \label{lhs3a} \ee

Now using (\ref{sum1}-\ref{sum2a}) we combine
all pairs of $I_+$ and $I_-$ in (\ref{lhs3a})
and then obtain

$$
\ln{(1-{1\over p})}+\ln{(1-{1\over p}e^{{\it i}\pi/3})}+
\ln{(1+{z\over p}e^{{\it i}\pi/3})}+\ln{(1-{z\over p}e^{{\it i}\pi/6})}
$$
which is the rhs of the relation (\ref{rel3}) after extracting the terms
$\ln p$.
For the domain IV (\ref{domain3}) where the fourth term in the rhs of (\ref{fes})
dominates we have the similar to (\ref{rel3}) relation
\be
{\ov u(p)}+u(p)+u(-\om p)-u(\om p)=
\ln{[(p-q')(p+\om^{-1}q')(p-\om q)(p-q)]}
\label{rel4}
\ee
which can be checked in the same way as the relation (\ref{rel3}).

The second functional relation (\ref{fes1}) can be proved similarly.
So we have checked that the functional relations (\ref{fes}) and (\ref{fes1})
are compatible with our result obtained from the two-line hypothesis.
Therefore, this fact can be considered as a justification of this hypothesis.

\section{Inversion relations}

In this section we want to check a compatibility of
 our solution (\ref{kappa4}) with the inversion relations.
Considering the product of two transfer matrices
$T_p\ov T_{p'}$ one can see that this product is decomposable
in two cases. Namely, one can fix
the parameters as follows $p'=p$ and $p'=-p$.
It is easy to see that the star weight defined by the formula (\ref{star0})
satisfies the following property:
\be
W^{star}(\a,\b,\g,\a;p,p,q,q')=
\Phi_{pq'}\d_{\b\g},\quad
W^{star}(\a,\b,\b,\d;p,-p,q,q')=
\Phi_{pq}\d_{\a\d}
\label{raz}
\ee
where the inversion factor
\be
\Phi_{pq}={{4(p^3+q^3)}\over{(p+q)^3}}.
\label{Phi}
\ee
Due to the property (\ref{raz}) one has the following decompositions of the
product of the transfer matrices
\be
T_p \ov T_{p} = \Phi_{pq'}^N\,(I + T^{(1)}),\quad
T_p \ov T_{-p} = \Phi_{pq}^N\,(I + T^{(2)})
\label{razval}
\ee
where $I$ is the identity matrix of the dimension $4^N$ and
$T^{(1)}$ ($T^{(2)}$) is a transfer matrix built up from some
weights $W^{(1)}$ ($W^{(2)}$).
Unfortunately, at the moment we can not establish rigorously
in what domain of the parameter
$p,q,q'$ the terms $T^{(1)}$ and $T^{(2)}$
in the rhs of equations (\ref{razval}) can be neglected
when $N\rightarrow\infty$.
So we want to
consider  how the inversion
relations work for the answer (\ref{kappa4}).

 Let us start with the expression
for the eigenvalues $t(p)$ and $\ov t(p)$ in the thermodynamic limit.
Using the results of section 7 we get
\be
\lim_{N\rightarrow\infty}{1\over N}\ln t(p) = \ln 2 + j(v_3^{-1})+j(v_4)+
\a+{{2i\pi}\over 3},\quad
\lim_{N\rightarrow\infty}{1\over N}\ln \ov t(p') = \ln 2 + j(v_1)+j(v_2^{-1})-
\a-{{2i\pi}\over 3}
\label{tptp'}
\ee
where $p,p',q,q'$ now are given by (\ref{qq'}) and
\be
j(v) = -{{i\pi}\over 9} -\ln(1+v)+I(v)
\label{j}
\ee
with $I$ given by (\ref{I}). The function
$\a=\lim_{N\rightarrow\infty} 1/N\log(a_{2N}(q,q'))$ with $a_{2N}$
entering to (\ref{sts}) is unknown but it cancels in the sum $\ln t(p)+
\ln \ov t(p')$. It is evident, that the function
$F$ defined by (\ref{FF}) is
\be
F(v)={1\over 4}\ln{{1+v}\over{1-v}}+{1\over 3}j(v)
\label{FJ}
\ee
and the logarithm of the partition function given by (\ref{kappa4})
can be also represented as
\be
\ln(\kappa/2)=K_1+K_2+K_3+K_4+{1\over 3}\chi(v_1,v_2,v_3,v_4)
\label{ka}
\ee
where
\be
\chi(v_1,v_2,v_3,v_4)=j(v_1)+j(v_2^{-1})+j(v_3^{-1})+j(v_4).
\label{xi}
\ee
It can be checked using the formulae (\ref{sum1}-\ref{sum2a}) that
\be
j(v)+j(-v^{-1})=\cases{
{\ds \;\;{{2i\pi}\over 3}} &  if
${\ds -{{5\pi}\over 6}<\arg v< -{{\pi}\over 6}}$\cr
\quad\cr
{\ds -{{2i\pi}\over 3}} &  if
${\ds \;\;{{\pi}\over 3}<\arg v< {{2\pi}\over 3}}$\cr
}
\label{inverJ1}
\ee
\be
j(v)+j(v^{-1})= \ln {{1+v^3}\over{(1+v)^3}} +
\cases{
{\ds -{{2i\pi}\over 3}} & if
${\ds -{{\pi}\over 6}<\arg v< {{\pi}\over 6}}$\cr
\quad\cr
{\ds -{{4i\pi}\over 3}} & if
${\ds \;\;{{2\pi}\over 3}<\arg v<\pi}$\cr
\quad\cr
{\ds \;\;{{2i\pi}\over 3} }& if
${\ds -\pi<\arg v< -{{2\pi}\over 3}}$\cr
}
\label{inverJ2}
\ee
Let us remind that in the domains I and II defined by (\ref{domain1})
and (\ref{domain2}) the inversion relations (\ref{rel1}) and (\ref{rel2}) are valid
for $s(p)$ and $\ov s(p)$ and therefore for $t(p)$ and $\ov t(p)$.
Another way to prove them is to use relations (\ref{inverJ1}-\ref{inverJ2}).
Indeed, if $p'=p$ then $1/v_3=q'/p, v_4=p, v_1=-1/p, 1/v_2=p/q'$ and
using (\ref{tptp'}) with the relations (\ref{inverJ1},\ref{inverJ2}) we get
$$
\lim_{N\rightarrow\infty}{1\over N}\ln t(p)+
\lim_{N\rightarrow\infty}{1\over N}\ln \ov t(p)=\ln 4+j(p)+j(-1/p)+j(p/q')+j(q'/p)=
$$

$$
\ln 4 +
\cases{
{\ds
-{{2i\pi}\over 3}+\ln{{p^3+{q'}^3}\over{(p+q')^3}} + {{2i\pi}\over 3}}& for
${\ds {{\pi}\over 3}<\arg p< {{\pi}\over 2}}$\cr
\quad\cr
{\ds
-{{2i\pi}\over 3}+\ln{{p^3+{q'}^3}\over{(p+q')^3}} - {{4i\pi}\over 3}}& for
${\ds {{\pi}\over 2}<\arg p< {{2\pi}\over 3}}$\cr
\quad\cr
{\ds
{{2i\pi}\over 3}+\ln{{p^3+{q'}^3}\over{(p+q')^3}} - {{2i\pi}\over 3}}& for
${\ds -{{2\pi}\over 3}<\arg p< -{{\pi}\over 3}}$\cr
}
=
$$
\be
\ln\Phi_{pq'}+\cases{
- 2i\pi & for ${\ds {{\pi}\over 2}<\arg p< {{2\pi}\over 3}}$\cr
\;\;0 & for ${\ds \>\>\>\> {{\pi}\over 3}<\arg p< {{\pi}\over 2}}
\>\>\>\>\> \mbox{and} \>\> {\ds \>\> -{{2\pi}\over 3}<\arg p< -{{\pi}\over 3}}$ \cr
}
\label{invpp}
\ee
i.e. up to $2i\pi$ the inversion relation (\ref{rel2}) is valid in the domain II.

For the second case $p'=-p$ we have $1/v_3=q'/p, v_4=p, v_1=1/p, 1/v_2=-p/q'$ and
we have
$$
\lim_{N\rightarrow\infty}{1\over N}\ln t(p)+
\lim_{N\rightarrow\infty}{1\over N}\ln \ov t(-p)=\ln 4+j(p)+j(1/p)+j(p/q')+j(-q'/p)=
$$
$$
\ln 4 +
\cases{
{\ds \ln{{p^3+1}\over{(p+1)^3}} - {{2i\pi}\over 3} + {{2i\pi}\over 3}}& for
${\ds {\ds -{{\pi}\over 6}<\arg p< {{\pi}\over 6}}}$\cr
\quad\cr
{\ds \ln{{p^3+1}\over{(p+1)^3}} - {{4i\pi}\over 3} - {{2i\pi}\over 3}}& for
${\ds {{5\pi}\over 6}<\arg p< \pi}$\cr
\quad\cr
{\ds \ln{{p^3+1}\over{(p+1)^3}} + {{2i\pi}\over 3} - {{2i\pi}\over 3}}& for
${\ds -\pi<\arg p< -{{5\pi}\over 3}}$\cr
}
=
$$
\be
\ln\Phi_{pq}+\cases{
-2i\pi & for ${\ds {{5\pi}\over 6}<\arg p< \pi}$\cr
\;\;0 & for ${\ds -{{\pi}\over 6}<\arg p< {{\pi}\over 6}} \>\>\>\>
\mbox{and} \>\>\> {\ds -\pi<\arg p< -{{5\pi}\over 3}}$ \cr
}
\label{invpm}
\ee
i.e. again the inversion relation (\ref{rel1}) is valid in the domain I up to $2i\pi$.

As we show below the inversion relation can also be written for the partition
function.
If \\
$2\pi/3<a_3<\pi$ then for the arguments of parameters $v_i$ given by (\ref{v})
one has
\beq
&{\ds -{{2\pi}\over 3}<\arg(v_1)<-{{\pi}\over 2},\quad
-{{\pi}\over 2}<\arg(-v_1^{-1})<-{{\pi}\over 3}}&\nonumber\\
&{\ds -{{2\pi}\over 3}<\arg(v_3^{-1})<-{{\pi}\over 2},\quad
-{{\pi}\over 2}<\arg(-v_3)<-{{\pi}\over 3}}&\nonumber\\
\label{v1v3}
\eeq
and
\beq
&{\ds -{{\pi}\over 6}<\arg(v_2)<0,\quad
0<\arg(v_2^{-1})<{{\pi}\over 6}}&\nonumber\\
&{\ds -{{\pi}\over 6}<\arg(v_4^{-1})<0,\quad
0<\arg(v_4)<{{\pi}\over 6}}&
\label{v2v4}
\eeq

So since (\ref{v1v3}) is valid one can use (\ref{inverJ1}) with
$v=v_1$,  $v=-v_1^{-1}$, $v=v_3^{-1}$ or $v=-v_3$, namely,
\be
j(v_1)+j(-v_1^{-1})={{2i\pi}\over 3},\quad
j(v_3^{-1})+j(-v_3)={{2i\pi}\over 3}.
\label{Jv1v3}
\ee
Similarly, due to (\ref{v2v4}) the relation (\ref{inverJ2})
is satisfied for  $v=v_2$,  $v=v_2^{-1}$, $v=v_4$ , $v=v_4^{-1}$
and one has
\be
j(v_2)+j(v_2^{-1})=\ln{{{p'}^3+{q'}^3}\over{(p'+q')^3}}-{{2i\pi}\over 3},
\quad
j(v_4)+j(v_4^{-1})=\ln{{{p}^3+{q}^3}\over{(p+q)^3}}-{{2i\pi}\over 3}.
\label{Jv2v4}
\ee
Since, the logarithm of the partition function (\ref{ka}) is connected
with the function $j$ via the combination $\chi$ let us show
the inversion relation for the function $\chi$:
\be
\chi(v_1,v_2,v_3,v_4)+\chi(-v_1^{-1},v_2^{-1},-v_3^{-1},v_4^{-1})=
\ln{{{p}^3+{q}^3}\over{(p+q)^3}}+\ln{{{p'}^3+{q'}^3}\over{(p'+q')^3}}.
\label{invxi}
\ee
So in the rhs of (\ref{invxi}) we have got the sum of
logarithms of the function which
is proportional to the product of the inversion factor  (\ref{Phi}).

Let us note that the function $j(-{\it i}e^{-{\it i}x})$ resembles the Baxter's
function $\hat J_3(x)$ (\ref{J_nhat}) because it satisfies the relations
(\ref{inverJ1}) and (\ref{inverJ2}) which are similar to the relations (10.1, 10.2)
of \cite{Bax} \footnote{The relations (10.1, 10.2) of \cite{Bax} were
written for the function $J_n$ (see (\ref{J_n})). To get such equations for
the function $\hat J_n$ one should take into account an additional term
$1/4\ln{\tan(x/2+\pi/4)}$ in (10.5) of \cite{Bax}}
A difference is that the relations
(\ref{inverJ1}-\ref{inverJ2}) are defined only in the restricted region
and have additional pure imaginary terms in the rhs
which are connected with the cuts of the function
$I$ entering to (\ref{j}). However, as we have shown above the relations
(\ref{inverJ1}-\ref{inverJ2}) are enough for the validity of the inversion relation
(\ref{invxi}). The function $\hat J_3(x)$ is analytic in a wider strip
$|\mbox{Re}(x)|<\pi/2$.
Of course, it also ensures the validity of the inversion relation (\ref{invxi}).
However,
the applicability domain for the inversion relation ensured by the  function
$\hat J_3(x)$ is wider. Namely, it works for the region $0<a_3<\pi$ while
our result satisfies the inversion relation when $2\pi/3<a_3<\pi$ as
was discussed above.

\section{The one-line regime}

Now let us consider the second regime II which differs from the regime I
by a transformation $a_i\rightarrow -a_i$. As was discussed above
the parameters $v_i$ should be substituted by the parameters $v'_i$ defined
by (\ref{vv}).

The numerical analysis shows that the transformation $a_i\rightarrow -a_i$
changes the ground state drastically in comparison with the regime I
where the two-line solution corresponds to the dominating eigenvalues
of the transfer matrices. Now the one-line solution corresponds to the
dominating regime.

 Namely, for the one-line solution
the imaginary parts $y_i$ defined by the formula (\ref{xiyi}) are close to $7\pi/12$
for all $i=1,\ldots,N$. As in the two-line case a precision of the approximation
$y_i-7\pi/12\approx 0$ becomes better with a growth of $N$. So, we also can accept
 the one-line hypothesis that in the thermodynamic limit $N\rightarrow\infty$ the
line corresponding to $y_i=7\pi/12$ tends to be exact.
It is interesting to note that now it is not necessary
to consider only those $N$ which are  multiples of $3$ as it was in the two-line
case. The one-line solution works for any $N$.

Now we would like to generalize the result of the sections 5 and 6 to the one-line
case. Somehow this case is simpler in comparison with the two-line case because
now we are dealing only with one line of the roots.
Therefore, it is not necessary to divide
the interval $1,\ldots,N$ into two parts as we did it in the section 5.
It is not very difficult to repeat here a strategy of the section 5, i.e. to
introduce a distribution density $\rho$ of the real parts $x_i$ in
the thermodynamic limit and
to obtain the integral equation for it. As a result we have the equation
\beq
&{\ds\ln{\Biggl[-
{
{\cosh{(k - {\it i} \pi/4)}\;-\;\cosh{(\zeta\;-\;{\it i}\pi/12)}}
\over
{\cosh{(k - {\it i} \pi/4)}\;-\;\cosh{(\zeta-{\it i}5\pi/12)}}
}
\Biggr]}+{\it i}\pi\;(1-2\int_{-\infty}^kdk'\rho(k'))}&\nonumber\\
&{\ds
- \int_{-\infty}^{\infty}dk'\rho(k')
\ln{\Biggl[{
{\cosh{(k - {\it i}\pi/4)}\;-\;\cosh{(k' +{\it i}5\pi/12)}}
\over
{\cosh{(k - {\it i}\pi/4)}\;+\;\cosh{(k' +{\it i}5\pi/12)}}
}
\Biggr]}
\;=\;0}&
\label{bethe1string}
\eeq
with the normalization condition
\be
\int_{-\infty}^{\infty}dk\rho(k)=1.
\label{normarho}
\ee

As the integral equations from the section 5 the equation (\ref{bethe1string})
can also be solved exactly.
The result looks as follows:
\be
\rho(k)={{\sqrt{3}/\pi}\over{2\cosh{[2(\zeta+k)]}+ 1}}+
{{\sqrt{3}/\pi}\over{2\cosh{[2(\zeta-k)]}- 1}}.
\label{rhoa}
\ee
Actually, one can rewrite this result as
\be
\rho(k)=\rho_+(-k)+\rho_-(k)
\label{rhob}
\ee
where $\rho_{\pm}$ are the densities in the two-line case
given by (\ref{resrho}). It is quite amazing that the density in the
one-line regime appears to be a superposition of the densities
for the two-line case.

Now we can calculate the logarithm of the partition function per site $\kappa$
which is still given by the formula (\ref{kappa3}) of section 6.
For the one-line regime $s_1$ and $s_2$ which were
introduced by (\ref{S1}) and (\ref{S2}) for the two-line regime
 are defined by
$$
s_1=
\int_{-\infty}^{\infty}dk\rho(k)\ln{(p+e^{\zeta-{\it i}\pi/4}
e^{k+{\it i}7\pi/12})}+
\int_{-\infty}^{\infty}dk\rho(k)\ln{(p+e^{\zeta-{\it i}\pi/4}
e^{-k-{\it i}7\pi/12})}
$$
\be
s_2=
\int_{-\infty}^{\infty}dk\rho(k)\ln{(p'-e^{\zeta+{\it i}\pi/4}
e^{k-{\it i}7\pi/12})}+
\int_{-\infty}^{\infty}dk\rho(k)\ln{(p'-e^{\zeta+{\it i}\pi/4}
e^{-k+{\it i}7\pi/12})}
\label{S12a}
\ee
where $p$ and $p'$ are still given by formula (\ref{qq'}) but for parameters
$v_i$ substituted by $v'_i$ defined above.
Using the solution (\ref{rhob})  and taking into account
(\ref{qq'}) we get $s_1$ and $s_2$ in the thermodynamic limit
\beq
&{\ds
s_1=
{{\sqrt{3}}\over{\pi}}\int_{-\infty}^{\infty}dk
{{\ln{[{\ds
-{{v'_4}\over{v'_3}}e^{-{\it i}\pi/6}\,(1-{v'_3}e^{k+i\pi/6})}]}}\over{2\cosh{2k}-1}}
+
{{\sqrt{3}}\over{\pi}}\int_{-\infty}^{\infty}dk
{{\ln{[{\ds
{{v'_4}\over{v'_3}}e^{-{\it i}\pi/3}\,(1+{v'_3}e^{k+i\pi/3})}]}}\over{2\cosh{2k}+1}}
}&\nonumber\\
\nonumber\\
&{\ds+
{{\sqrt{3}}\over{\pi}}\int_{-\infty}^{\infty}dk
{{\ln{[{\ds
v'_4\,(1- {v'_4}^{-1} e^{k+i\pi/6})}]}}\over{2\cosh{2k}-1}}
+
{{\sqrt{3}}\over{\pi}}\int_{-\infty}^{\infty}dk
{{\ln{[{\ds v'_4\,(1+{v'_4}^{-1}e^{k+i\pi/3})}]}}\over{2\cosh{2k}+1}}}&
\label{S1b}
\eeq

\vspace{0.4cm}

\beq
&
s_2=
{\ds
{{\sqrt{3}}\over{\pi}}\int_{-\infty}^{\infty}dk
{{\ln{[{\ds -{1\over{v'_1}}\,(1-{v'_2}e^{k+i\pi/6})}]}}\over{2\cosh{2k}-1}}
+
{{\sqrt{3}}\over{\pi}}\int_{-\infty}^{\infty}dk
{{\ln{[{\ds
-{1\over{v'_1}}\,(1+{v'_2}e^{k+i\pi/3})}]}}\over{2\cosh{2k}+1}}
}&\nonumber\\
\nonumber\\
&{\ds+
{{\sqrt{3}}\over{\pi}}\int_{-\infty}^{\infty}dk
{{\ln{[{\ds
e^{-{\it i}\pi/6}\,(1- {v'_1}^{-1} e^{k+i\pi/6})}]}}\over{2\cosh{2k}-1}}
+
{{\sqrt{3}}\over{\pi}}\int_{-\infty}^{\infty}dk
{{\ln{[{\ds
-e^{-{\it i}\pi/3}\,(1 + {v'_1}^{-1} e^{k+i\pi/3})}]}}\over{2\cosh{2k}+1}}.
}&
\label{S2b}
\eeq

\vspace{0.1cm}
After some algebra we come
to the following result for the logarithm of the partition function in the one-line case:
\be
\ln{(\kappa/2)}=F'({v'_1}^{-1})+F'(v'_2)+F'(v'_3)+F'({v'_4}^{-1})
\label{kappa1string}
\ee
where the function $F'(v)$ differs from the function $F(v)$ defined
by the formulae (\ref{FF}-\ref{Ipm}) only by the pure imaginary term:
\be
F'(v)={{5{\it i}\pi}\over{216}}-
{1\over{12}}\ln{(1+v)}-{1\over{4}}\ln{(1-v)}+{1\over 3}I(v).
\label{FF'}
\ee
One can check that, as in formula  (\ref{kappa4}),
the pure imaginary term compensates the imaginary part coming from other terms
making the final answer (\ref{kappa1string}) real.

Let us note that proceeding as in the section 7
we can also check that the one-line solution is compatible with the
functional relations in the thermodynamic limit.

\section{Conclusion}

Let us summarize the basic results obtained above.
Our main point here was to calculate the partition function
of the Zamolodchikov model for $n=3$ layers based on the functional relations
and the Bethe ansatz equations obtained in our previous works
\cite{BM1},\cite{BM2}. We have considered two regimes I and II.
Analysing numerically the structure of
the solutions to the Bethe ansatz equations corresponding to the
largest eigenvalues we have come to the two-line hypothesis for
the regime I and one-line hypothesis for the regime II.
Using the standard technique
we have got the integral equations
for the distribution densities
in the thermodynamic limit
which are solvable
exactly due to a difference property of the kernel of these integral
equations. We have compared the results obtained from the integral
equations with the discrete distributions for the finite size of the lattice.
A similarity of the discrete distributions with the corresponding
solutions of the integral equations pretending to be the continuous
limits of the discrete distributions tells us about a plausibility
of our conjectures.

We have used these distribution densities for a calculation of the partition
function. The answer can be expressed in terms of  the dilogarithm
function. It is in a good agreement with the data obtained
from the numeric solutions to the Bethe ansatz equations up to $N=90$.
We have also checked a compatibility of this result with the
functional relations.

It would be very interesting  to apply the above
calculations to the three-layer Zamolodchikov model without a modification
of the boundary conditions.
We hope to do that in our further publication. We also hope that proceeding
this way we could get a better understanding of the
problem discussed above and to study in thermodynamic limit
the ground state of the hamiltonian derived by Baxter and Quispel
in paper \cite{BQ}.

We also intend to develope the thermodynamic Bethe ansatz (TBA) technique
for a study of the finite size corrections and possible conformal
properties of the three-layer Zamolodchikov model.

As a further step we would like to generilize our results obtained for the
three-layer case to a generic case of arbitrary number of layers.

\section{Acknowledgements}

The authors would like to thank M.~Batchelor,
R.~Baxter, V.~Bazhanov, L.D.~Faddeev, R.~Flume,
G.~von~Gehlen, M.~Karowski, J-M.~Maillet and V.~Rittenberg
for stimulating discussions and suggestions.
This research (HB) has been  supported by the Alexander von Humboldt Foundation
and (VM) by the Australian Research Council. HB would also like to thank
R.~Schrader for his kind invitation to Berlin Free University where the work
on this paper was continued.

\end{document}